\documentclass[11pt]{article}

\usepackage{arxiv}

\usepackage[utf8]{inputenc} 
\usepackage[T1]{fontenc}    
\usepackage{url}            
\usepackage{booktabs}       
\usepackage{amsthm,amsmath,amsfonts}
\usepackage{natbib}
\usepackage{float}
\usepackage{subcaption}
\usepackage{algorithm2e}
\usepackage{xcolor}
\usepackage{multirow}
\usepackage{color,soul}
\usepackage{blkarray}
\usepackage{nicefrac}   
\usepackage[colorlinks,citecolor=blue,urlcolor=blue,filecolor=blue,backref=page]{hyperref}
\usepackage{microtype}      
\usepackage{lipsum}
\usepackage{graphicx}
\usepackage{anyfontsize}
\usepackage{afterpage}
\usepackage{adjustbox}
\usepackage{color, colortbl}
\usepackage{mathbbol}
\usepackage{mathrsfs}
\usepackage{hhline}
\usepackage{rotating}
\usepackage{comment}
\usepackage{tabularx}
\usepackage{sectsty}

\definecolor{Gray}{gray}{0.9}

\allowdisplaybreaks

\usepackage{arydshln}
\makeatletter
\renewcommand*\env@matrix[1][*\c@MaxMatrixCols c]{%
	\hskip -\arraycolsep
	\let\@ifnextchar\new@ifnextchar
	\array{#1}}
\makeatother

\RequirePackage{doi}

\renewcommand{\tilde}[1]{\widetilde{#1}}

\usepackage{bbm} 

\newcommand{\bbE}{\mathbb{E}}
\newcommand{\bbV}{\mathbb{V}}

\newcommand{\lrnd}{\left(}
\newcommand{\rrnd}{\right)}
\newcommand{\lsq}{\left[}
\newcommand{\rsq}{\right]}
\newcommand{\lcur}{\left\lbrace}
\newcommand{\rcur}{\right\rbrace}
\newcommand{\given}{\,|\,}

\newcommand{\tCollapsedurl}{\if0\blind
	{
		\url{\tCollapsedurl}
	} \fi
	\if1\blind
	{ 
		\url{github.com/}\texttt{[url redacted in blinded version]}
	} \fi}

\title{Finite mixtures in capture-recapture surveys for modelling residency patterns in marine wildlife populations}

\author{
Gianmarco Caruso\\
    \scriptsize{Dpt. of Statistical Sciences}\\
    \scriptsize{Sapienza University of Rome}\\
    \scriptsize{\texttt{gianmarco.caruso@uniroma1.it}}
\And 
    Pierfrancesco Alaimo Di Loro\\
    \scriptsize{Dpt. GEPLI}\\
    \scriptsize{LUMSA}\\
    \scriptsize{\texttt{p.alaimodiloro@lumsa.it}}
\And
    Marco Mingione\\
    \scriptsize{Dpt. of Political Sciences}\\
    \scriptsize{Roma Tre University}\\
    \scriptsize{\texttt{marco.mingione@uniroma3.it}}
\And
    Luca Tardella\\
    \scriptsize{Dpt. of Statistical Sciences}\\
    \scriptsize{Sapienza University of Rome}\\
    \scriptsize{\texttt{luca.tardella@uniroma1.it}}
\And
    Daniela Silvia Pace\\
    \scriptsize{Dpt. of Environmental Biology}\\
    \scriptsize{Sapienza University of Rome}\\
    \scriptsize{\texttt{danielasilvia.pace@uniroma1.it}}
\And
   Giovanna Jona Lasinio\\
    \scriptsize{Dpt. of Statistical Sciences}\\
    \scriptsize{Sapienza University of Rome}\\
    \scriptsize{\texttt{giovanna.jonalasinio@uniroma1.it}}
}

\begin{document}

\RestyleAlgo{boxruled}
	
	\def\spacingset#1{\renewcommand{\baselinestretch}%
		{#1}\small\normalsize} \spacingset{1}
		
\maketitle
\begin{abstract}
This work aims to show how prior knowledge about the structure of a heterogeneous animal population can be leveraged to improve on its abundance estimation from capture-recapture survey data. We combine the Open Jolly-Seber (JS) model with finite mixtures and propose a parsimonious specification tailored to the residency patterns of the common bottlenose dolphin. We employ a Bayesian framework for our inference, discussing the appropriate choice of priors to mitigate label-switching and non-identifiability issues, commonly associated with finite mixture models. We conduct a series of simulation experiments to illustrate the competitive advantage of our proposal over less specific alternatives. 
The proposed approach is applied to data collected on the common bottlenose dolphin population inhabiting the Tiber River estuary (Mediterranean Sea). Our results provide novel insights into this population's size and structure, shedding light on some of the ecological processes governing its dynamics. 
\end{abstract}

\newpage
\section{Introduction} \label{sec:intro}

Capture-recapture (CR) methods are statistical techniques widely employed to estimate the size of an elusive population for which it is impossible to get a complete enumeration. This task, applied initially to ecology for the study of fish and wildlife populations \citep{otis1978statistical,wu2017bayesian,matechou2022capture}, is now common to many other application fields such as epidemiology \citep{chao2001applications,BOHNING2020197,maruotti2023estimating} and social sciences \citep{bohning2009recent,brittain2009estimators,bohning2018capture,silverman2020multiple,di2020bayesian,farcomeni2022many}.
The term \textit{capture} is inherited from the traditional way wild animals have been identified for decades - namely through capture, marking and release - but it is not necessarily intended for its physical sense anymore. Researchers increasingly adopt non-invasive methods for monitoring wild populations to minimise the costs and impact on the population of interest. Among those, photo-identification  
\citep{royle2009bayesian,pace2021capitoline} and DNA sampling \citep{bravington2016close,morin2016monitoring} are becoming increasingly popular as they minimise behavioural responses that may bias the final estimates \citep[see][and references therein]{fegatelli2013improved}.

Original applications of such methods date back to the beginning of the 20th century and were based on standard homogeneity assumptions on the population structure and the identification process \citep{le1965note, amstrup2010handbook}.
The literature is now rich in alternatives that can address a large variety of deviations from such basic model assumptions and suit situations where, for example, individuals exhibit heterogeneous behaviours \citep{pledger2000unified}, sampling occurs in continuous time \citep{altieri2022continuous}, stop-over sites are present \citep{matechou2013integrated,worthington2019estimating,wu2021bayesian}, temporary emigration is allowed \citep{zhou2019removal}, and so on.
For an exhaustive review, see \cite{KING201933} and references therein. 

Our work concentrates on the abundance estimation of a \textit{common bottlenose dolphin} population inhabiting a delimited area over multiple years. Such population is known to be open and calls for using the Open Jolly-Seber model, a standard CR framework for open populations \citep{amstrup2010handbook}.  
Furthermore, the established ecological literature affirms that it comprises individuals with different residency patterns and results in a population clustered into groups with different levels of site-fidelity. 
Hence, the homogeneity assumption of the standard CR framework cannot hold, and the behaviour of individuals belonging to different groups (i.e. entrance, capture and survival) must be described by different parameters. Accurate estimation of the clustering structure and parameters is of utmost interest to ecologists who want to describe the population's dynamics and inform conservation policies.
A widespread practice in such a heterogeneous setting is to consider the inclusion of individual covariates that can help explain the differences among the population's members. However, informative covariates are often unavailable and the heterogeneity is entirely latent, as it is in the case study under consideration. 
Finite Mixture Models (FMM) represent the natural solution to this impasse.  Each individual is assigned to a different mixture component with its own set of common and distinct parameter values. FMM approaches to CR models have been successfully employed in a likelihood-based framework both in closed population \citep{pledger2000unified,dorazio2003mixture,pledger2005performance} and open population \citep{pledger2010open, guery2017hidden} settings. 
From the Bayesian perspective, attempts have been made to model heterogeneity in detection and behavioural effects in closed populations by \cite{ghosh2005bayesian}. More recently, \cite{turek2021bayesian} proposed a non-parametric finite mixture model with an unknown number of components and different capture probabilities.\\
We build on the \cite{pledger2003open,pledger2010open}'s FMM extension to the Open JS model to account for latent heterogeneity. 
We embed this finite mixture approach in the \cite{royle2008hierarchical,royle2012parameter}'s parameter-expanded data-augmentation (PX-DA) framework, which turns out to be particularly convenient to fit Bayesian CR models via standard Markov Chain Monte Carlo (MCMC) algorithms. We discuss its implementation challenges and introduce suitable prior specifications that mitigate the label-switching and non-identifiability issues of FMM. 
The model is tested through an extended simulation study and applied to photo-identification CR survey data of the common bottlenose dolphins (\emph{tursiops truncatus}) population inhabiting the area of the Tiber River estuary in the Mediterranean Sea.
In particular, we show how the prior scientific knowledge on the population of interest can be leveraged to specify a parsimonious FMM tailored to its supposed structure and check its validity. When this is the case, we show it can sensibly improve the performances of more comprehensive specifications.  

The remainder of the paper is organised as follows: Section \ref{sec:Motivating} describes the motivation and intuitions behind our work, with a brief description of the data that will be considered later on; Section \ref{sec:JSmodel} illustrates a Bayesian hierarchical formulation of the \cite{pledger2003open,pledger2010open}'s Jolly-Seber (JS) class of mixture models within the \cite{royle2008hierarchical,royle2012parameter}'s data-augmentation framework and introduces our modelling proposal as a parsimonious alternative; Section \ref{sec:simstudy} reports a simulation experiment to investigate different aspects of the model and its estimation; Section \ref{sec:data_analysis} provides the model choice and results on the set of data originally introduced as the motivating example.

\section{Motivating example}\label{sec:Motivating}
\begin{figure}[ht]
    \centering
     \begin{subfigure}[b]{.48\textwidth}
       \includegraphics[width=.95\textwidth]{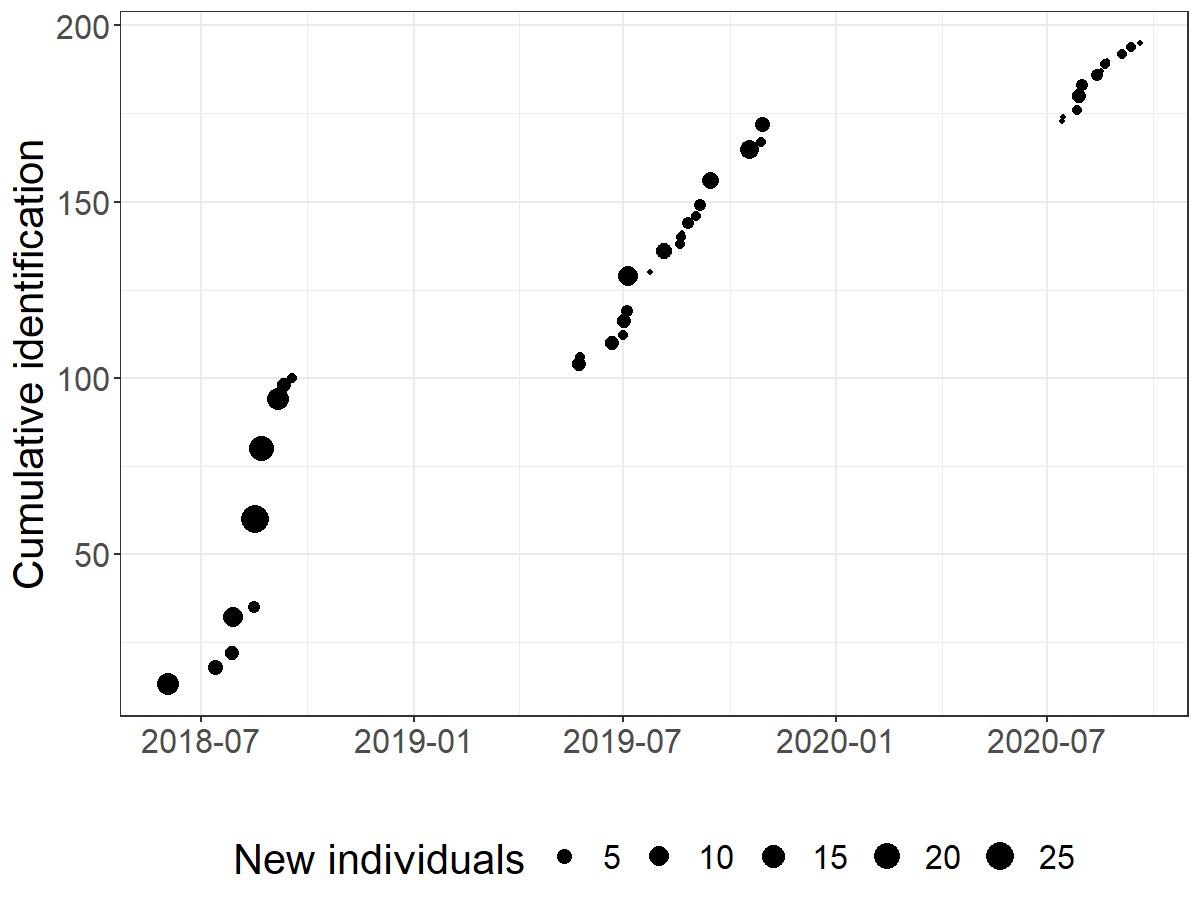}
       \caption{}
       \label{cumdisc}
   \end{subfigure}
   \begin{subfigure}[b]{.48\textwidth}
       \includegraphics[width=.95\textwidth]{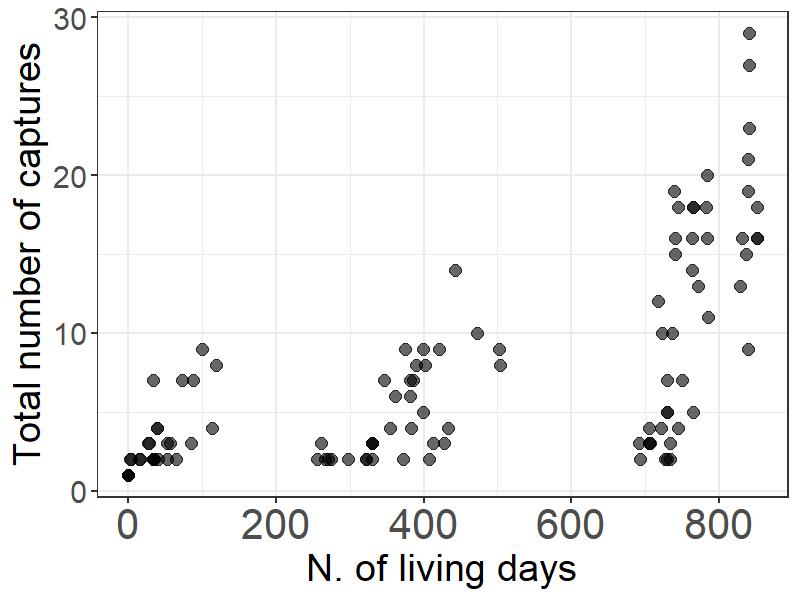}
       \caption{}
       \label{caphist}
   \end{subfigure}
    \caption{(a) Cumulative number of individual identifications with size proportional to the number of newly identified individuals; (b) Total number of captures by individual.}
    \label{fig:datadesc}
\end{figure}

This work is motivated by the need to get an abundance estimate of the population of the common bottlenose dolphins inhabiting the area of the Tiber River estuary in the Mediterranean Sea.
Boat-based daily surveys have been conducted between 2018 and 2020 under favourable weather conditions to collect photographic and acoustic data of the specimens encountered in the study area during the search \citep{papale2021higher, pace2022bray, pace2022resources, pace2022seasonal}. The photo-identification technique was used to identify unique individuals over multiple sampling occasions and to build single capture histories. For this analysis, we focus on the so-called \textit{well-marked} individuals since the probability of their misidentification can be assumed to be negligible\footnote{Photo-identification is based on matching the same marks across different pictures; the level of the mark (poor, fair, well) is an index of how identifiable a dolphin is.}. 
As a consequence, the final estimates are related to the subset of well-marked individuals only, representing a portion of the population visiting the study area.
Figure \ref{cumdisc} shows the cumulative number of identified individuals across the different sampling occasions, where the size of each point is proportional to the number of newly identified individuals. Its trend is known as the \textit{discovery rate}, which is maximum during the first year (notice that 50\% of new identifications occurred in 2018, with a maximum of 25 newly identified individuals registered in August) and slowly decreases over time. Further details about the study area, the data collection process and the analysis are available in \cite{pace2021capitoline}. \\
Most of the recent literature about common bottlenose dolphins converges towards the identification of three groups characterised by different levels of site-fidelity to a specific area: from the most to the least frequently present \citep{dinis2016bottlenose, hunt2017demographic, haughey2020photographic, la2022determinants}.
This feature is of utter interest to biologists interested in disentangling the permanent or semi-permanent population from the transient one.  Generally speaking, one group is composed of individuals that (almost) never leave the study area; these are usually referred to as \textit{resident} individuals, observable on many occasions and for long periods of time, and expected to have the largest number of captures (see Figure \ref{caphist}). The other group includes individuals who are not continuously present in the study area but regularly visit it; these are called \textit{part-time} individuals, observable throughout a wide time window but usually encountered at occasions far apart in time. The last group comprises individuals that enter the study area only once in their lifetime for a short time window and whose captures are rare; these are \textit{transient} individuals that are observable only on occasions occurring on close dates.
This inherent heterogeneity cannot be neglected without hindering the reliability of the final abundance estimates \citep{gimenez2018individual}. 
To this end, \cite{pace2021capitoline} applies a hierarchical clustering and includes the group labels in the POPAN-JS framework \citep{schwarz1996general} to model heterogeneity in the entrance, capture and survival probabilities.
They find signs of the expected structure and estimate significant differences in most components of the JS model. While practical for understanding the underlying structure of the population of interest, this two-step approach has some relevant issues. First of all, there is neither quantification nor propagation of the uncertainty of the classification step onto the modelling step; this can bias the final estimates and yield over-confident conclusions. Second, the same information set is used twice in two different statistical procedures, where the latter is performed after conditioning on the first; this can lead to a confirmation bias in favour of the original hypothesis.

This paper proposes to unify the two steps into a joint statistical procedure. We embed FMM into the Open-JS framework in a Bayesian hierarchical setting, allowing for the estimation of cluster labels and population size altogether while properly propagating the uncertainty at all levels \citep{clark2006hierarchical}.

\section{State of the art and proposed model}\label{sec:JSmodel}
Let $D$ be the number of distinct individuals observed at least once during $T$ distinct sampling occasions. Data are collected in a ($D\times T$) matrix, say $\mathbf{Y}$, with the generic element recording whether individual $i=1,\dots,D$ individual has been observed ($y_{it}=1$) or not ($y_{it}=0$) at occasion $t=1,\dots,T$. Hence, the rows of $\mathbf{Y}$ contain the capture histories of all the encountered individuals.
The open-population JS model \citep{schwarz1996general} envisions individuals entering (i.e., via birth or immigration) and exiting (i.e., via death or emigration) the population during the sampling occasions. 
The emigration is assumed to be permanent for identifiability purposes, i.e. once an individual has left the population, it cannot return to it. 
Furthermore, the JS models assume that all captures are independent across individuals and over time.
The latter is achieved by considering the existence of a super-population of unknown size, say $N_{super}$, accounting for all the individuals potentially available (encountered or not) in the study area between the first and the last sampling period.  The super-population size is the main parameter of interest and determines the dimension of the model parameter space. Its typical estimation in the Bayesian framework involves jumping between spaces of different dimensions, hence requiring the implementation of Reversible Jump MCMC algorithms \citep{brooks2000bayesian}. Here, we adopt the data-augmentation approach proposed by \cite{royle2008hierarchical} to keep the dimension of the parameter space \textit{fixed} throughout the iterations, thus bypassing the need for the Reversible Jump MCMC \citep{arnold2010capture}.

\subsection{Parameter Expanded Data-Augmentation (PX-DA) formalisation}
\label{subsec:JSintro}
The Parameter Expanded Data-Augmentation (PX-DA) approach by \cite{royle2008hierarchical} sets the dimension of the parameter space equal to a fixed value $M\gg N_{super}\geq D$. 
In this way, solving the varying dimension issue by converting it into a more manageable \textit{missing data} problem in the multi-state processes context (see Section \ref{multistate} of the Appendix). The rows of the observed data matrix $\mathbf{Y}$ are augmented to $M$, hence defining $\mathbf{Y}_{\text{aug}}=\{\mathbf{Y},\boldsymbol{0}_{M-D}\}$, where $\boldsymbol{0}_{M-D}$ is an $(M-D)\times T$ matrix of zeroes. 
$M$ must be set such that $N_{super}\in\left\lbrace D, D+1, \dots, M\right\rbrace$ and, consequently, $N_{super}-D$ among the $M-D$ rows of zeroes correspond to individuals who belong to the super-population but have never been encountered. 
The remaining $M-N_{super}$ correspond to \textit{pseudo}-individuals that have never been part of the population during the observation window and hence do not belong to the super-population.
The data generation process assumes that an individual can be recruited into the population at the beginning of each sampling period only if it has never been recruited on previous occasions. On the other hand, individuals who have already been recruited can leave the population between two subsequent sampling periods. 
All the $D$ observed individuals will eventually be recruited in the population between the first and the last capture occasion, as they have been captured at least once. 
This dynamic is controlled through two time-varying latent binary variables: the first one, $r_{it}$, is equal to $1$ if and only if individual $i$ is recruitable at time $t$ ($0$, otherwise); the second one, $z_{it}$, is equal to $1$ if and only if individual $i$ belongs to the population at time $t$ ($0$, otherwise).

All $M$ individuals are recruitable at the first occasion (i.e. $r_{i1}=1$, $\,i = 1, \dots, M$), while they become permanently non-recruitable once they have entered the population.
Let $\rho_t,\, t=1,\dots,T$ be the recruitment probabilities, i.e. the probability that an available (not yet entered) individual in the augmented dataset is recruited in the population at time $t$. Let $\phi_t,\, t=2,\dots, T$, be the apparent survival probability\footnote{The term \textit{apparent} is included as permanent emigration and mortality are indistinguishable and, thus, treated as the same phenomenon in JS-type models.}, i.e. the probability that a recruited individual is in the population at the following sampling occasion.  
The following rules govern the latent label process: $z_{i1}\sim Bern(\rho_1),\, i = 1, \dots, M$, \, $r_{it}=\text{min}\{r_{i,t-1},1-z_{i,t-1}\}$ and $z_{it}\,|\, z_{i,t-1},r_{it}\sim Bern(\phi_t\cdot z_{i,t-1}+\rho_t\cdot r_{it})$, $t=2,\dots,T$.
Thus, the distribution of the generic element of the augmented data matrix can be expressed conditionally on $z_{it}$ as:
\begin{equation}
\label{eq:obs_process}
    y_{it}|z_{it}\sim Bern(p_t\cdot z_{it})\,,
\end{equation}
where $p_t$ is the capture probability at occasion $t$.
Since $y_{it}=0$ almost surely when $z_{it}=0$, \eqref{eq:obs_process} is a zero-inflated binomial model.
Notice that, given $M$, the marginal likelihood of the capture histories can be expressed as:
\begin{equation}
\label{eq:lik1}
\mathcal{L}_M(\boldsymbol{y}\,|\,\boldsymbol{\theta})=\int_{\mathcal{Z}}\prod_{i=1}^M\prod_{t=1}^T p(y_{it}\,|\,z_{it})\cdot p(z_{it}\,|\,z_{i,t-1}, r_{it})\, p(dz_{it}),
\end{equation}
where $\mathcal{Z}\subset \lbrace 0,1\rbrace^{M\times T}$ contains all possible $z_{it}$ configurations and $\boldsymbol{\theta}$ is the full set of model parameters $\boldsymbol{\theta}=\lcur p_t, \phi_t, \rho_t\rcur_{t=1}^T$.
The integral in the likelihood expression has no general closed-form solution and numerically marginalising out the latent components  can be computationally intensive. This problem is common to the Hidden Markov Models (HMMs) literature, in which some efficient techniques based on the forward algorithm have been proposed to efficiently solve the integration problem \cite{worthington2019estimating}. Alternatively, MCMC algorithms provide a viable solution to work, iteration after iteration, with the conditional likelihood:
\begin{equation*}
    \mathcal{L}^c_M(\boldsymbol{y}\,|\,\boldsymbol{\theta}, \boldsymbol{z})=\prod_{i=1}^M\prod_{t=1}^T p(y_{it}\,|\,z_{it}),
\end{equation*}
and approximate the posterior distribution of all the quantities of interests in a Bayesian setting that provides full uncertainty quantification. Furthermore, standard MCMC methods are now relatively easy to implement by practitioners due to the availability of software like JAGS
\citep{plummer2003jags} or NIMBLE \citep{de2017programming}.

Marginally, the hierarchical model specification resulting from Equation \eqref{eq:obs_process} implies that $N_{super}\sim Binom(M,\,\psi)$, where $\psi$ is the overall inclusion probability (throughout all occasions) of any individual in the super-population.
\cite{royle2008hierarchical} show that $\psi$ is linked to the recruitment probabilities through the following equation:
\begin{equation}
    \label{eq:inclusion_prob}
    \psi=1-\prod_{t=1}^T\,(1-\rho_t).
\end{equation}
In particular, \eqref{eq:inclusion_prob} implies that $\mathbb{E}[N_{super}\,|\, M,\rho_1,\dots,\rho_T]=M\,\Bigl[1-\prod_{t=1}^T\,(1-\rho_t)\Bigr]=M\,\psi$. Hence, choosing the prior distribution for $\rho_1,\dots,\rho_T$ is crucial to determining the prior on $N_{super}$. \cite{dorazio2020objective} demonstrates that the prior:
\begin{equation}
    \label{eq:rhoPrior}
    \rho_t\sim Beta\biggl(\frac{1}{T},\,2-\frac{t}{T}\biggr), \, \quad t = 1, \dots, T,
\end{equation}
induces an objective prior on $N_{super}$.
In terms of practical inference, the estimated population size at each time $t$ and the overall super-population size can be derived through the latent variables $z$'s, namely $N_t=\sum_{i=1}^M\,z_{it}$ and $N_{{super}}=\sum_{i=1}^M\,\mathbb{1}_{\{\sum_{t=1}^T\,z_{it}>0\}}$. 

\subsection{Jolly-Seber (JS) finite mixture modelling for open populations}
\label{subsec:FMM}
Most populations are composed of individuals with heterogeneous behaviours. In some cases, the heterogeneity can be assumed to be well-described by a finite number (say $G$) of different patterns. These depend on the individual's latent traits, and their consideration requires complex modelling tools, such as FMM.
The underlying assumption of FMM is that each unit can belong to only one group $g=1,\dots, G$, with unknown prior probabilities $w_g$ ($\sum_gw_g=1$). The individuals belonging to different groups in open population CR studies may have different capture, recruitment, or survival parameters. In the most general specification, the relative order among the parameters of different groups can change at each time $t$ (e.g. group one could have the highest detection rate at the first sampling occasion but the lowest at the second one). This JS-type mixture model is known as the Interactive Heterogeneous Model (IHM) and its very rich specification depends on too many parameters for successful model fitting \citep[see][for further discussion]{pledger2010open}.
\cite{pledger2003open, pledger2010open} explore simpler specifications that could adequately represent the population structure and introduce a convenient notation to navigate through all possible sub-models. Let $t$ and $h$ be the time and group heterogeneity effects, respectively. Different expressions correspond to different modelling structures: constant in time and homogeneous across groups $(\cdot)$; time-varying but homogeneous across groups $(t)$; constant in time but heterogeneous across groups $(h)$; time-varying and heterogeneous across groups with separable interaction $(t+h)$; time-varying and heterogeneous across group with non-separable interaction $(t\times h)$. For example, the IHM corresponds to $\left\lbrace \left[\rho_{t\times h},\, \phi_{t\times h},\, p_{t\times h}\right]_G \right\rbrace$, where the underlying population is supposed to be composed by $G$ classes. If we want to specify a model whose heterogeneous group effect lies in the capture probabilities only, we write $\{\left[{\rho}_{t},\, {\phi}_{t},\, {p}_{t\times h_G}\right]\}$. The subscript is moved to highlight that the mixture of $G$ components is related only to detection. We will take advantage of this notation in the sequel of the paper.

\subsection{Modelling class heterogeneity using finite mixtures within the PX-DA approach}
\label{subsec:mixture_models}
We embed the PX-DA formalisation of the open JS model into FMM by adding one layer of hierarchy in the original hierarchical specification. 

Now, let $c_i\in\lbrace 1,\dots, G\rbrace$ be the latent membership label of each individual $i = 1, \dots,M$ in $\mathbf{Y_{aug}}$. The full hierarchical specification is as follows:
\begin{equation}
\label{eq:hiermod_gen}
\begin{aligned}
    &y_{it}\,|\, z_{it},c_i=g\sim Bern(p_{gt}\cdot z_{it}),\\
    &z_{it}\,|\, z_{i,t-1},r_{it},c_i=g\sim Bern(\phi_{gt}\cdot z_{i,t-1}+\rho_{gt}\cdot r_{it}), \quad  r_{it}=\text{min}\{r_{i,t-1},1-z_{i,t-1}\},\\
    & p_{gt}\sim \pi_{p_g}(\cdot), \quad \phi_{gt}\sim \pi_{\phi_g}(\cdot), \quad \rho_{gt}\sim \pi_{\rho_g}(\cdot), \quad c_i\sim Multinom\left(1, (w_1,\dots, w_G)\right),\\
    & (w_1,\dots, w_G)\sim \pi_{\mathbf{w}}(\cdot)\,
\end{aligned}
\end{equation}
where $\pi_.(\cdot)$ refers to a generic prior distribution for the parameter.
The hierarchical formulation of \eqref{eq:hiermod_gen} further complicates the likelihood expression of Equation \eqref{eq:lik1} to:
$$
\mathcal{L}_M(\boldsymbol{y}\,|\,\tilde{\boldsymbol{\theta}})=\sum_{g=1}^G w_g\cdot\int_{\mathcal{Z}}\prod_{i=1}^M\prod_{t=1}^T p_g(y_{it}\,|\,z_{it})\cdot p_g(z_{it}\,|\,z_{i,t-1}, r_{it})dz_{it},,
$$
where $\tilde{\boldsymbol{\theta}}=\lcur\textbf{w}, \lcur\boldsymbol{\theta}_g\rcur_{g=1}^G\rcur$ includes the prior weights of the cluster components and the component-specific sets of parameters. Its evaluation needs further marginalisation with respect to the latent group labels, and, once again, MCMC methods are a viable solution to achieve Bayesian estimation of all quantities of interest.

This modelling framework includes many possible specifications, according to what varies with time and across groups. We further generalise the model specification by adapting the survival mechanism to describe not equally spaced capture occasions. Indeed, the assumption of constant survival across identical time scales does not transfer to the not equally spaced scenario. When this is the case - i.e. $l_t=(\tau_t-\tau_{t-1}), t=2,\dots, T$ are the time differences between subsequent occasions - the survival probabilities should be appropriately compounded. Once the time scale is set (e.g., days, weeks, months, years, etc.), we have $\phi_{gt}=\phi_g^{l_t}$, where $\phi_g$ represents the survival probability across a single time-unit on the chosen scale. In addition, following \cite{pledger2003open}, a convenient and parsimonious way to express the time-varying capture probabilities is through the logit link, i.e. $\text{logit}(p_{gt})=\mu_g+\tau_t\,,\, t=1,\dots,T\,,$ where $\mu_g$ determines the overall average capture probability of each group and $\tau_t$ is an occasion-specific differential effect.

The last step of the Bayesian model specification involves the choice of prior distributions for all parameters and their hyper-parameters. The natural prior for the mixture weights in finite mixture models is the Dirichlet distribution $Dir_G(\alpha_1,\dots,\alpha_G)$. It corresponds to a uniform distribution over the $G$-dimensional simplex when $\alpha_g=1,\;\forall g$, x is usually adopted as weakly informative prior.
For $\rho_{gt}, t=1,\dots, T$, we follow \cite{dorazio2020objective} and use the prior in \eqref{eq:rhoPrior}. 
General-purpose and weakly informative priors can be ascribed to $\mu_g$. At the same time, we suggest considering a $N(0,\,\sigma^2)$ on each parameter $\tau_t$, with $\sigma^2$ small, to induce only small time variations (in the logit-scale) on the capture probabilities. We further impose that $\sum_t\tau_t= 0$ to favour the otherwise weak identifiability of the $\mu_g$ and the $\tau$'s \citep{pledger2003open}.
Nevertheless, such choices do not defend the model from the label-switching problem of the class-specific parameters typical of the FMM framework because of the likelihood invariance under permutations of the components' labels.

\paragraph{Choosing priors on class-specific parameters}\label{sec:priors}
The choice of the prior distribution on component-specific parameters of FMMs can be seen as a challenge and an opportunity. This class of models suffers from several sources of non-identifiability. In our setting, the most affected are recruitment and survival probabilities. However, the final inference on the population size is usually robust with respect to their proper identification \citep[see][and references therein]{mena2015bayesian}. The recruitment probabilities $\rho_{gt}$ are model devices that allow the entrance of new individuals in the population, accounting for the openness of the population \citep{royle2008hierarchical}. They can be treated as nuisance parameters without a solid biological meaning and their identifiability is not of great concern. On the other hand, the capture and survival probabilities are of biological interest. They can be seen as indicators of how long and how often units of different types visits the sampling area. 

A weakly informative or regularising choice of their prior distributions can favour identifying these components. In turn, this can ease the identification of the membership labels already affected by the well-known \textit{label-switching} problem. 
Among the different solutions that have been proposed in the literature, one relies on imposing an ordering constraint among the component-specific parameters, i.e. $u_1<\dots<u_G$ \citep{richardson1997bayesian, diebolt1994estimation, chung2004difficulties}. This task is trivial in the context of parameters belonging to the whole real domain, where truncating or shifting Gaussian distributions allows full control of the position and scale of the priors. The latter is a reasonable choice for the $\mu_g,\,g=1,\dots,G$ parameters. 
However, when $u_g\in\left(0, 1\right),\;\forall\,g$ as in the case of the survival probabilities, the conditional prior specification must account for their bounded domains.
The standard solution would consider conditionally specified Uniform distributions. However, we exploit the alternative proposal of \cite{alaimodiloro2022} that allows for more flexibility. It is based on the Beta distribution and its generalisations (\textit{truncated} or \textit{restricted}), and it is both effective in imposing the constraint and controlling for the shape and first moments of the induced marginal priors. In other words, one may set $u_1\sim Beta(\alpha_1,\,\beta_1)$ and 
\begin{equation}
    \begin{aligned}
    & u_g|u_{g-1}\sim tBeta(\alpha_g,\,\beta_g; u_{g-1},1), \, \text{ or } 
    & u_g|u_{g-1}\sim rBeta(\alpha_g,\,\beta_g; u_{g-1},1)
    \end{aligned}\qquad g=2,\dots, G,
\end{equation}
where \textit{tBeta} and \textit{rBeta} represent the Truncated Beta and Restricted Beta, respectively, and it is possible to derive closed-form expressions for the marginal prior distribution of $u_g, g\geq 2$, or at least its first and second moment. Further computational details are given in Section \ref{app:betapriors} of the Appendix, along with formal proofs of the original results.

\subsection{The RPT model for the common bottlenose dolphin population}
\label{sec:botdolModel}
The population structure illustrated in Section \ref{sec:Motivating} can be translated to reflect variations in the parameters of the JS-type PX-DA framework. We have that the resident (R) individuals, showing high site-fidelity, should be characterised by high survival and capture probabilities; the part-time resident (P) individuals, with average site-fidelity, should be characterised by high survival probabilities but eventually undetectable on some occasions; the transient individuals (T) only shortly visit the study area and hence should have very low survival.
We want to build a suitably flexible encompassing model such that it is possible to establish the amount of evidence in favour of this assumption. In other words, we shall not enforce this exact structure in our model, but we can make it so that it is recognisable if present.
For instance, we could allow the three groups to have group-specific survival and capture probabilities as in the full specification of \eqref{eq:hiermod_gen}. That might be a too flexible model with a huge number of parameters, therefore yielding highly uncertain estimates because of weak identifiability. Alternatively, we can propose a more parsimonious specification tailored to the supposed behavioural differences among individuals from different groups. In this way, some components can be assumed to be common to different groups, reducing the model's complexity and favouring its identifiability.

First of all, we model the population recruitment dynamic assuming that each group is characterised by its own set of time-varying recruitability parameters, i.e. $\rho_{R,\,t},\,\rho_{P,\,t},\,\rho_{T,\,t}$. 
This setting naturally induces a clustering in the relationship between the recruitment process and the super-population size $N_{super}$, yielding group-specific inflation parameters that express the total population as the sum of three sub-populations. This modifies the analytical expression of the expected super-population size as follows:
\begin{equation*}
    \bbE\lsq N_{super}|\psi_{R},\psi_{P},\psi_{T},w_{R},w_{P},w_{T}\rsq=M\cdot\sum_{g=R,P,T}\,w_g \psi_{g} \,,
\end{equation*}
where $w_g$ and $\psi_g$ ($g=R,P,T$) are, respectively, the $g$-th component-specific mixture weight and inflation parameter.

Second, we separate the three sub-populations into two distinct classes, characterised by different tendencies to stay in the study area.  We can distinguish between short-term survivors, i.e. transient individuals who visit the area for narrow time windows, and long-term survivors, i.e. non-transient individuals (resident or part-time) who visit the area for wide time windows. To model such behavioural heterogeneity, we assume that one group (T, transient) has a smaller survival probability ($\phi_{T}$) than the other two groups ($\phi_{NT}$): $\phi_{T}<\phi_{NT}$.
Notice that these two parameters represent the survival probabilities across a single time unit on the chosen scale. When the capture occasions are not equally spaced, the survival probabilities should be appropriately compounded along different lengths (cfr. Section \ref{subsec:mixture_models}). In practice, given the two base survival probabilities $\phi_{T}$ and $\phi_{NT}$, the actual survival probabilities at time $t$ are:
$$
\phi_{T,\,t} = \phi_{T}^{l_t},\quad \phi_{NT,\,t} = \phi_{NT}^{l_t},
$$
where $l_t=|t - t'|,\, t=1,\dots,T$ are the time lags between the $T$ subsequent occasions.
Finally, the proposed model retains the time-varying structure of the detectability, as already introduced in Section \ref{subsec:mixture_models}, but discriminates between part-time (P) and non-part-time (NP) individuals introducing a \textit{partial undetectability} component in the latter group. Loosely speaking, part-time individuals are allowed to be \textit{undetectable} while \textit{alive} (because temporarily not present in the study area) on some occasions chosen at random with probability $\delta\in(0,1)$.
This corresponds to modelling the capture probability on occasion $t$ as:
\begin{equation*}
\begin{aligned}
    & p_{NP,\,t}=\text{logit}^{-1}\lrnd \mu+\tau_t\rrnd \, \text{ and } 
    & p_{P,\,t}=(1-\delta)\,\cdot\,p_{NP,\,t},
\end{aligned}\quad\,  t=1,\dots,T,
\end{equation*}
for the NP and P individuals, respectively; of course, $p_{P,\,t}<p_{NP,\,t},\,\forall t$. Notice that the larger $\delta$ is, the more the part-time group is separated from the resident group.
This parameter plays a similar role to the completely random emigration parameter of \cite{kendall1997estimating} and it is needed to control the temporary emigration pattern of \textit{part-time} individuals discussed in Section \ref{sec:Motivating}. Appendix \ref{app:tempem} shows that such parametrisation is indeed equivalent to temporary emigration under the simplifying assumption of emigration occurring at random. We name this model RPT as it encompasses three specific types of behaviour. However, we would like to point out how the model does not enforce this interpretation. For instance, both survivals could be estimated to be high, or the undetectability parameter could be estimated as approximately equal to $1$, etc.

\section{Simulation experiments}\label{sec:simstudy}
We conduct a simulation experiment to assess the performances of the RPT model whenever it is well-specified, i.e. the data are generated according to the structure described in Section \ref{sec:botdolModel}. 
We generate multiple sets of artificial data under alternative scenarios from the RPT with fixed parameters and then estimate a pool of models on them. Our main objective is twofold: i) to evaluate the ability to recover the true values of the parameters, with a particular focus on $N_{super}$; ii) to verify whether the RPT is chosen as the \textit{best} among other alternatives based on some model selection criterion. 

We consider four scenarios characterised by an increasing number of sampling occasions, i.e. $T\in\lcur 10, 20, 30, 40\rcur$, to verify the model performances for different time horizons. We suppose that in the first scenario (i.e. $T=10$), all the captures are recorded within a relatively short period (e.g. within a year). Longer time horizons are included in the other scenarios, where a larger time gap (year gap) is assumed to occur every $10$ occasions. Further details about the time lags are available in Section \ref{app:timelags} of the Appendix.
The month (and portion of months) is taken as the basic time unit to avoid the possible numerical instability related to the large values of the lags in terms of days. Note that this affects the interpretation of the survival probability parameter as it must be interpreted as the probability of surviving one month.
We adopt the following parameters' values in all the scenarios. The survival probabilities are set to $\phi_T=0.01$ and $\phi_{NT}=0.997$. These two values may seem quite extreme at first glance. However, they guarantee that the short-time survivors (transient individuals) almost surely stay in the population for less than a year and that the long-term survivors (non-transient individuals) stay in the population for more than three months with a very high probability ($>0.99$). Furthermore, notice that a monthly survival probability equal to $0.01$ corresponds to a survival probability equal to $0.86$ on a daily scale and equal to $0.34$ on a weekly scale. On the other hand, a monthly survival probability equal to $0.997$ corresponds to a probability of $0.87$ on a four-years scale. Therefore, the monthly scale appears as a good compromise to avoid a value that is too low for $\phi_{T}$ and a value that is too high for $\phi_{NT}$. The capture probabilities are obtained by setting $\mu=0$ and $\delta=0.7$, and generating $\tau_t\sim N(0, 0.25)$ in each scenario. The recruitment parameter for transient individuals is fixed to $\rho_{T,\,t}= 0.02$ for all capture occasions (assuming a negligible temporal variation for this group). At the same time, we consider the following structure for resident and part-time individuals: $$\rho_{R,\,t}=\begin{cases} 0.4\,,\quad t=1\\0.02\,,\quad t=10k+1\,,\,k=1,2,3\\0.0025\,,\quad\text{otherwise} \end{cases}\qquad\text{and}\qquad \rho_{P,\,t}=\begin{cases} 0.4\,,\quad t=1\\0.04\,,\quad t=10k+1\,,\,k=1,2,3\\0.005\,,\quad\text{otherwise} \end{cases}.$$ 
The mixture weights for the three groups are set to $w_R=0.2$, $w_P=0.45$ and $w_T=0.35$.
We envision an augmented super-population of $M^*=500$, that yields an expected super-population size $\bbE\lsq N_{super} \rsq \in \lbrace170, 209, 243, 271\rbrace$ for $T=10, 20, 30, 40$, respectively. Notice that $N_{super}$ increases with $T$ as more individuals can visit the study area during a longer time horizon.

We simulate independent encounter histories for $K=50$ pseudo-populations. 
Along with the RPT model, we consider 10 different alternatives in the class of JS-type models described in Section \ref{subsec:FMM} and \ref{subsec:mixture_models}, all having time-varying recruitment parameters. In the simplest case, we consider a model with homogeneous capture and survival probability (M$_1$: $\{\rho_t,\,\phi,\,p_t\}$). When the population is supposed to be structured in $G$ groups, we suppose that each group has its own time-varying recruitment probability and that mixture components may vary by capture probability (M$_2$-M$_4$: $\{\rho_{t\times h_G},\,\phi,\,p_{t+h_G}\}$, for $G=2,3,4$), by survival probability (M$_5$-M$_7$: $\{\rho_{t\times h_G},\,\phi_{h_G},\,p_t\}$, for $G=2,3,4$) or by both (M$_8$-M$_{10}$: $\{[\rho_{t\times h_G},\,\phi_{h},\,p_{t+h}]_G\}$, for $G=2,3,4$). 
Each simulated dataset was augmented by $500$ all-zero capture histories to implement the PX-DA approach so that $D_k+500=M_k\neq M^*$ for all simulated sets. 
We consider the prior setting described in Section \ref{sec:JSmodel}. 
Notably, we specify a $N(0,\,10)$ for the intercept $\mu$. When the survival probability is the same for all the individuals, a standard Uniform prior is placed on that parameter. When two survival probabilities are considered (generically, $\phi_1$ and $\phi_2$), we use $\phi_1\sim Beta(1,\,2)$ and $\phi_2|\phi_T \sim tBeta(1,\,1,\, \phi_T,\,1)$, with the latter marginally yielding $\phi_2\sim Beta(2,\,1)$. Then enforcing $\phi_1<\phi_2$ induces a slight repulsion between the two parameters. When one considers more than two survival parameters, constrained Uniform priors are instead chosen. These different prior choices for the models do not substantially affect the model selection criterion associated with each model but - conversely - are useful to induce a better separation between couples of survival parameters. Estimation is carried out using JAGS, in which we run $2$ chains with $20,000$ iterations each, discarding $5,000$ as burn-in and thinning by $2$ the remainder to save storage space \citep{brooks2004bayesian}.

We chose median posterior estimates instead of simple averages, mitigating the effect of anomalies that can result in occasionally low-informative datasets. In the same spirit, we rely on the \textit{Mean Absolute Error} (MAE) as an accuracy measure instead of the widely employed \textit{Root Mean Squared Error}. Interval estimation is assessed through the percentage of times the $95\%$ credible intervals contain the true value of the parameter (i.e. the coverage) and the average $95\%$ credible interval width (CIW). The overall goodness-of-fit is measured via the Watanabe-Akaike Information Criterion (WAIC, \cite{watanabe2010asymptotic}), following the good practice of \cite{gelman2014understanding} whenever finite mixtures are fitted. We also report the overlapping index ($OV$, \cite{pastore2018overlapping, pastore2019measuring}) between the posterior distributions of the two survival parameters, averaged over the $K = 50$ replicas. If $OV=0$, the two distributions are completely separated, while if $OV=1$, the two distributions perfectly overlap. This metric is particularly appealing to understand whether the posterior distributions of the group-specific parameters are well-separated or not, justifying the related model parameterisation. Finally, we investigate the fuzzy classification ability of the RPT model using multi-class AUC (mAUC) \citep{hand2001simple}. Let us remark that the classification performances can only be evaluated in the simulation setting and for the RPT model, as the true group labels are known and are consistent with the estimated ones. Notice that individuals who have not been observed are not provided with a capture history, thus, it is impossible to infer the group they belong to. Therefore, the estimated mixture weights (referred to as the full $M_k$-sized pseudo-population) are not expected to align with the mixture weights employed in the data generation process. Indeed, the latter was used to generate the cluster labels of the whole pseudo-population, of which some individuals (i.e. the pseudo-individuals) never become part of the real population.

\begin{figure}[ht]
    \centering
    \includegraphics[width=0.8\textwidth]{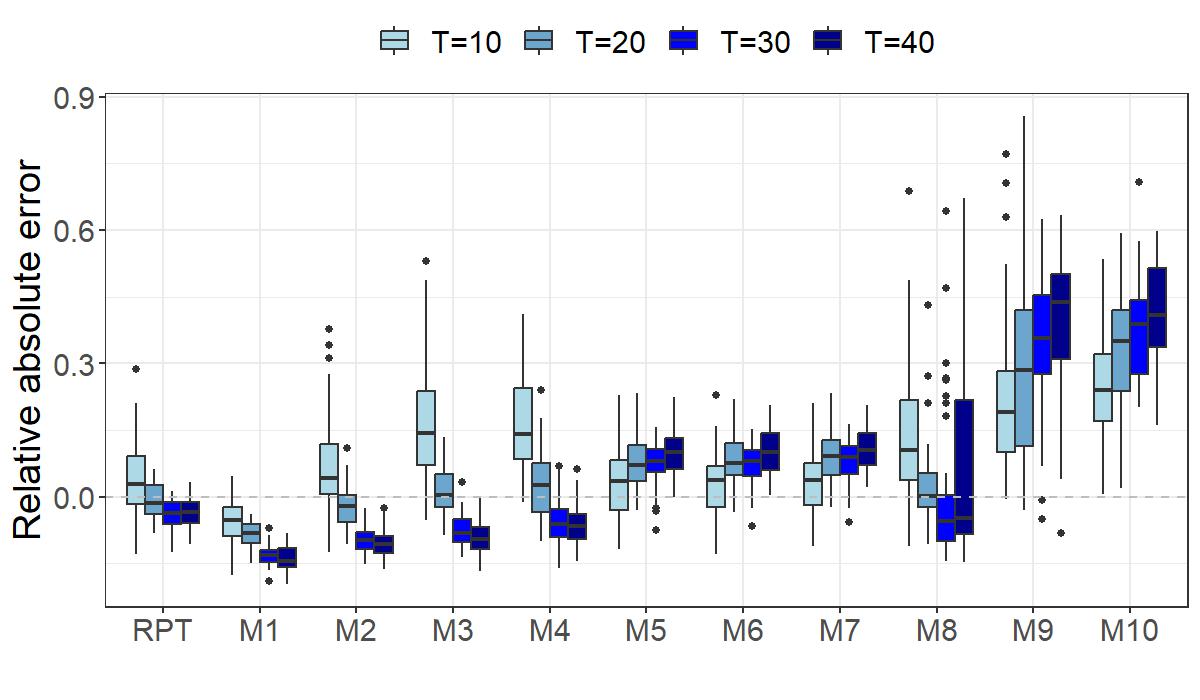}
    \caption{Relative estimation error of the super-population ($N_{super}$) abundance for increasing number of sampling occasions ($T$), calculated for each of the $K=50$ independent replicas, by the RPT and the ten alternative \cite{pledger2010open}'s mixture models.}
    \label{fig:boxplot_Nsuper}
\end{figure}

Figure \ref{fig:boxplot_Nsuper} shows the differences between the estimated ($\hat{N}_{{super}}$) and true ($N_{{super}}$) super-population size for each of the $K=50$ replicas. The error is divided by the expected value of $N_\text{super}$ in the corresponding scenario to allow for a meaningful comparison between scenarios having expected super-population sizes of different magnitudes.
We notice that the RPT uniformly provides the best results overall. On the contrary, its competing models consistently underestimate or overestimate the super-population size. Notably, as $T$ increases, the underestimation is more evident for those models that do not account for heterogeneity in survival probabilities (i.e. M$_1$-M$_4$) - cfr. \cite{pledger2003open,pledger2010open} -, while the overestimation is substantial for those that consider an excessive number of parameters (i.e. M$_8$-M$_{10}$). Intuitively, a greater number of parameters controlling the capture probabilities tends to infer an excessive number of uncaught individuals from the zero-histories.

Table \ref{tab:Nsup} reports useful summaries to assess the models' performances in estimating $N_\text{super}$. The MAE and CIW are scale-dependent measures and do not allow for scenario comparisons. Therefore, we consider relative versions of these measures by dividing both by the corresponding expected value of $N_\text{super}$. 
\begin{table}[ht]
    \centering
    \caption{Estimates of relative MAE (MAE$_\text{rel}$), coverage (Cov.) and relative average width of the $95\%$ credible intervals (CIW$_\text{rel}$) for $N_{super}$, median WAIC and percentage of times each competing model has achieved best WAIC (\%waic). All these summaries have been obtained when data are simulated from the RPT model.}
    \label{tab:Nsup}
     \resizebox{1\textwidth}{!}{%
\setlength\extrarowheight{5pt}
\begin{tabular}{l|ccccc|ccccc|ccccc|ccccc}
 \centering
 & \multicolumn{5}{c}{{{$T = 10$}}}
 & \multicolumn{5}{c}{{{$T = 20$}}}
 & \multicolumn{5}{c}{{{$T = 30$}}} 
 & \multicolumn{5}{c}{{{$T = 40$}}}\\
\cmidrule[0.3pt](l){2-6} 
\cmidrule[0.3pt](l){7-11} 
\cmidrule[0.3pt](l){12-16} 
\cmidrule[0.3pt](l){17-21} 
\textbf{Model} &
  \textbf{MAE$_\text{rel}$} &
  \textbf{Cov.} &
  \textbf{CIW$_\text{rel}$} & 
  \textbf{WAIC} & 
  \textbf{\%waic} &
  \textbf{MAE$_\text{rel}$} &
  \textbf{Cov.} &
  \textbf{CIW$_\text{rel}$} & 
  \textbf{WAIC} & 
  \textbf{\%waic} &
  \textbf{MAE$_\text{rel}$} &
  \textbf{Cov.} &
  \textbf{CIW$_\text{rel}$} & 
  \textbf{WAIC} &
  \textbf{\%waic} &
  \textbf{MAE$_\text{rel}$} &
  \textbf{Cov.} &
  \textbf{CIW$_\text{rel}$} & 
  \textbf{WAIC} &
  \textbf{\%waic}\\
  \midrule
  RPT & 0.07 & 0.96 & 0.28 & 1433.96 & 6 & 0.04 & 0.96 & 0.17 & 2953.01 & 50 & 0.04 & 0.86 & 0.15 & 4353.83 & 86 & 0.04 & 0.92 & 0.15 & 5840.58 & 88 \\
  $\text{M}_1:\,\left\lbrace \rho_t,\, \phi,\, p_t \right\rbrace$ & 0.06 & 0.70 & 0.15 & 1458.18 & 2 & 0.08 & 0.28 & 0.10 & 3175.72 & -- & 0.13 & 0 & 0.06 & 4875.18 & -- & 0.14 & 0 & 0.06 & 6668.57 & -- \\
  $\text{M}_2:\,\left\lbrace \rho_{t\times h_2},\, \phi,\, p_{t+h_2} \right\rbrace$ & 0.08 & 0.94 & 0.33 & 1448.27 & 8 & 0.05 & 0.92 & 0.16 & 2991.11 & -- & 0.10 & 0.10 & 0.09 & 4495.79 & -- & 0.11 & 0.24 & 0.10 & 6136.03 & -- \\
  $\text{M}_3:\,\left\lbrace \rho_{t\times h_3},\, \phi,\, p_{t+h_3} \right\rbrace$ & 0.15 & 0.88 & 0.69 & 1435.05 & 10 & 0.05 & 0.94 & 0.25 & 3012.28 & -- & 0.08 & 0.60 & 0.15 & 4532.3 & -- & 0.09 & 0.42 & 0.13 & 6133.42 & -- \\
  $\text{M}_4:\,\left\lbrace \rho_{t\times h_4},\, \phi,\, p_{t+h_4} \right\rbrace$ & 0.16 & 0.84 & 0.65 & 1439.58 & 32 & 0.06 & 0.92 & 0.28 & 2985.86 & -- & 0.07 & 0.70 & 0.18 & 4510.22 & -- & 0.08 & 0.64 & 0.17 & 6169.65 & -- \\
  $\text{M}_5:\,\left\lbrace \rho_{t\times h_2},\, \phi_{h_2},\, p_t \right\rbrace$ & 0.07 & 0.90 & 0.26 & 1448.64 & 14 & 0.08 & 0.90 & 0.27 & 3089.37 & 10 & 0.08 & 0.96 & 0.27 & 4710.57 & -- & 0.10 & 0.76 & 0.27 & 6436.28 & -- \\
  $\text{M}_6:\,\left\lbrace \rho_{t\times h_3},\, \phi_{h_3},\, p_t \right\rbrace$ & 0.06 & 0.90 & 0.25 & 1438.10 & 4 & 0.08 & 0.90 & 0.26 & 3071.13 & 18 & 0.08 & 0.96 & 0.26 & 4706.34 & 6 & 0.10 & 0.74 & 0.27 & 6426.04 & 8\\
  $\text{M}_7:\,\left\lbrace \rho_{t\times h_4},\, \phi_{h_4},\, p_t \right\rbrace$ & 0.06 & 0.92 & 0.26 & 1435.73 & 12 & 0.08 & 0.84 & 0.26 & 3094.53 & 22 & 0.09 & 0.92 & 0.27 & 4725.57 & 8 & 0.11 & 0.72 & 0.27 & 6452.76 & 4\\
  $\text{M}_8:\,\left\lbrace \left[\rho_{t\times h},\, \phi_{t\times h},\, p_{t\times h}\right]_2 \right\rbrace$ & 0.12 & 0.90 & 0.56 & 1440.04 & -- & 0.05 & 0.96 & 0.21 & 3041.38 & -- & 0.10 & 0.68 & 0.47 & 4668.08 & -- & 0.12 & 0.64 & 0.64 & 6453.97 & -- \\
  $\text{M}_9:\,\left\lbrace \left[\rho_{t\times h},\, \phi_{t\times h},\, p_{t\times h}\right]_3 \right\rbrace$ & 0.20 & 0.84 & 0.83 & 1430.92 & 4 & 0.28 & 0.62 & 0.76 & 2967.91 & -- & 0.36 & 0.40 & 0.74 & 4444.86 & -- & 0.41 & 0.20 & 0.66 & 6071.77 & -- \\
  $\text{M}_{10}:\,\left\lbrace \left[\rho_{t\times h},\, \phi_{t\times h},\, p_{t\times h}\right]_4 \right\rbrace$ & 0.25 & 0.78 & 0.84 & 1437.25 & 8 & 0.32 & 0.52 & 0.85 & 2951.84 & -- & 0.38 & 0.40 & 0.75 & 4424.14 & -- & 0.42 & 0.14 & 0.59 & 6055.66 & -- \\
  \bottomrule
\end{tabular}
}
\end{table}
The RPT model returns the lowest relative MAEs in all the scenarios that involve more than 1 year of observation, while its competitors are associated with errors, which mostly increase with $T$. The WAIC seems to fail in selecting the RPT model when the number of sampling occasions is rather small (i.e. $T = 10$) by attaining the lowest values in correspondence of models that do not account for survival heterogeneity (i.e. M$_1$-M$_4$ in $52\%$ of the replicas). This is a reasonable outcome since it is extremely complicated (if not impossible) to distinguish individuals with low and high survivals when the occasions span only 1 year. Notably, in such a short-term scenario, the low survival of a transient individual is likely to be confounded with an exceptionally low capture probability. This would also explain why model $\,\left\lbrace \rho_{t\times h_4},\, \phi,\, p_{t+h_4} \right\rbrace$ with four component-specific capture probabilities is selected in almost $1$ out of $3$ replicas according to the WAIC.
However, as $T$ increases, the WAIC tends to favour the true model, yielding the lowest median score and selecting it (i.e., returning the lowest WAIC) in most replicas.  
The classification performances of the RPT model are quite good in all the scenarios, with the mAUC improving as $T$ increases. Notably, the resulting median mAUC is always $\geq 0.82$ and above $0.95$ when a  year change occurs. Furthermore, by assigning the labels to each encountered individual according to the Maximum a Posteriori (MAP) rule, the median accuracy (across replicas) lies between $75\%$ and $95\%$ in all scenarios (once again improving as $T$ increases).

Table \ref{tab:paramest} shows the RPT model performance in estimating some time-constant parameters. The estimates of the undetectability and the survival parameters have a very small mean absolute error, although they fail to attain the nominal credible interval coverage of $95\%$ in most scenarios. This indicates a good accuracy of the point estimates associated with over-confidence (visible in the low average CIWs), thus reducing the nominal coverage. Nonetheless, it settles to a fair and acceptable level.
The result is not particularly surprising since these component-specific parameters govern latent ecological processes and are potentially mutually confounded.
\begin{table}[ht]
    \centering
    \caption{Estimates of MAE, coverage (Cov.) and average width of the $95\%$ credible intervals (CIW) for some time-constant parameters of model RPT. All these summaries have been obtained when data are simulated from the RPT model.}
    \label{tab:paramest}
     \resizebox{1\textwidth}{!}{%
\setlength\extrarowheight{5pt}
\begin{tabular}{ccccccccccccccccc}
 \centering
 & \multicolumn{3}{c}{{{$\phi_T$}}}
 & \multicolumn{3}{c}{{{$\phi_{NT}$}}}
 & \multicolumn{3}{c}{{{$\delta$}}}
 & \multicolumn{3}{c}{{{$\mu$}}} \\
\cmidrule[0.3pt](l){2-4} 
\cmidrule[0.3pt](l){5-7} 
\cmidrule[0.3pt](l){8-10} 
\cmidrule[0.3pt](l){11-13} 
\textbf{$T$} &
  \textbf{MAE} &
  \textbf{Cov.} &
  \textbf{CIW} & 
  \textbf{MAE} &
  \textbf{Cov.} &
  \textbf{CIW} & 
  \textbf{MAE} &
  \textbf{Cov.} &
  \textbf{CIW} & 
  \textbf{MAE} &
  \textbf{Cov.} &
  \textbf{CIW} \\
  \midrule
10 & 0.30 & 0.82 & 0.88 & 0.04 & 0.62 & 0.09 & 0.06 & 0.94 & 0.25 & 0.27 & 0.94 & 0.98 \\
20 & 0.08 & 0.88 & 0.35 & 0.005 & 0.68 & 0.01 & 0.03 & 0.82 & 0.10 & 0.13 & 0.82 & 0.43 \\
30 & 0.07 & 0.80 & 0.25 & 0.002 & 0.76 & 0.007 & 0.03 & 0.62 & 0.07 & 0.08 & 0.90 & 0.32 \\
40 & 0.06 & 0.82 & 0.20 & 0.001 & 0.80 & 0.05 & 0.02 & 0.74 & 0.06 & 0.07 & 0.96 & 0.27 \\
\bottomrule
\end{tabular}
}
\end{table}
Finally, we observe that the two survival probabilities (i.e. $\phi_T$ and $\phi_{NT}$) are well separated in all scenarios, with an $OV$ equal to $0.078$ when $T=10$ and equal to $0$ for larger $T$.

\section{Real data analysis} \label{sec:data_analysis}

We apply the RPT model to estimate the total population size of the common bottlenose dolphins inhabiting the Tiber River estuary, as introduced in Section \ref{sec:Motivating}. 
The data are the detection histories of $D = 195$ well-marked dolphins that have been sighted in the area between June 2018 and November 2020, for a total of  $T = 87$ occasions. 
After some preliminary runs with different values of $M$, we finally set $M-D = 500$ rows of pseudo-individuals (i.e. with null capture histories), thus yielding $M = 695$. Other choices of $M$ led to similar results, with larger values of $M$ only straining the computational burden, in terms of runtime and storage.

We run $2$ parallel chains, each with $20,000$ iterations with a burn-in of $5,000$ iterations and no thinning. We compare the performances of the RPT model with the wide range of alternative models illustrated in Section \ref{sec:simstudy}, using the same prior setting specified in the simulation study for all the considered models.
Table \ref{tab:realdata} reports the results on abundance estimation along with the WAIC associated with each competing model.
\begin{table}[ht]
    \centering
    \caption{WAIC, estimated total super-population abundance and by group ($95\%$ credible intervals).}
    \label{tab:realdata}
    
    \resizebox{.5\textwidth}{!}{%
    \setlength\extrarowheight{5pt}
    \begin{tabular}{ccc}
    \toprule
        \textbf{Model} &  \textbf{WAIC} & $\hat{N}_{super}$ \\
        \midrule
        RPT & 4725.1 & 311 (266, 373) \\
        $\left\lbrace \left[\rho_{t\times h},\, \phi_{t\times h},\, p_{t\times h}\right]_3 \right\rbrace$ & 4801.7 & 600 (467, 669) \\
        $\left\lbrace \left[\rho_{t\times h},\, \phi_{t\times h},\, p_{t\times h}\right]_4 \right\rbrace$ & 4811.9 & 594 (409, 664)\\
        $\left\lbrace \rho_{t\times h_2},\, \phi_{h_2},\, p_t \right\rbrace$ & 4844.6 & 341 (287, 418) \\
        $\left\lbrace \rho_{t\times h_3},\, \phi,\, p_{t+h_3} \right\rbrace$ & 4870.3 & 274 (243, 299) \\
        $\left\lbrace \rho_{t\times h_2},\, \phi,\, p_{t+h_2} \right\rbrace$ & 4877.9 & 276 (240, 307) \\
        $\left\lbrace \rho_{t\times h_4},\, \phi_{h_4},\, p_t \right\rbrace$ & 4893.1 & 352 (281, 458) \\
        $\left\lbrace \rho_{t\times h_4},\, \phi,\, p_{t+h_4} \right\rbrace$ & 4903.8 & 271 (244, 306) \\
        $\left\lbrace \rho_{t\times h_3},\, \phi_{h_3},\, p_t \right\rbrace$ & 4932.2 & 334 (269, 441) \\
        $\left\lbrace \left[\rho_{t\times h},\, \phi_{t\times h},\, p_{t\times h}\right]_2 \right\rbrace$ & 5088.5 & 356 (267, 628) \\
        $\left\lbrace \rho_t,\, \phi,\, p_t \right\rbrace$ & 5108.7 & 230 (216, 242) \\
        \bottomrule
    \end{tabular}
    }
\end{table}
We notice that the RPT model does yield the lowest WAIC score. Interestingly, the second best choice according to the WAIC is model $\left\lbrace \left[\rho_{t\times h},\, \phi_{t\times h},\, p_{t\times h}\right]_3 \right\rbrace$, which indeed resembles the same structure as the RPT model but without allowing for mixture components sharing common parameters. This lack of parsimony results in an overestimation of the individuals in the super-population, similar to what has been observed in the simulation study (cfr. Section \ref{sec:simstudy}). Thus, these results suggest the presence of unobserved heterogeneity in the population, which seems to be better described by the more parsimonious structure of the RPT model than by a generic three-group specification.

The two annual survival probabilities are well separated with an $OV = 0$ and posterior estimates are $\hat{\phi}_T = 1.06\times 10^{-8}$ ($CI_{0.95} = [0, 2.8\times 10^{-6}]$) for the group of transient individuals and $\hat{\phi}_{NT} = 0.71$ ($CI_{0.95} = [0.62, 0.80]$) for the resident and part-time individuals.
Notice that the estimated survival parameter $\hat\phi_T$ is of little interpretability on the annual scale. However, it corresponds to a probability of $0.73$ on the weekly scale and $0.26$ on the monthly scale. The average capture probability of the resident and transient individuals is $\hat{p}_{NP}=\text{logit}^{-1} (\hat{\mu})\approx 0.19$; the corresponding temporal variations, captured by $\tau_t$, result in the time-dependent posterior distributions of $p_{NP,\,t}$ reported in Figure \ref{fig:pt_real} in the Appendix. 

The parameter $\delta$ regulating the undetectability of the part-time individuals is estimated to be $\hat{\delta}\approx 0.74$. This means that individuals in that group are present in the area approximately for the $26\%$ of their lifetime.
Although the estimates of the recruitment parameters have little interpretation in the \cite{royle2008hierarchical}'s considered framework, it is worth noticing that, on average, the recruitment probabilities are higher during the first year for the resident individuals, while approximately constant for part-time and transient individuals (cfr. Figure \ref{fig:rhot_real} in the Appendix).
This is a model artefact motivated by all the individuals already present in the population before the start of the survey (i.e. mostly residents) and that are \textit{virtually} recruited on the first occasion. If we consider the second year only (most reliable in terms of recruitment probability estimation and individual classification), the average recruitment probabilities are approximately $7\times10^{-4}, 1\times 10^{-3}, 6\times 10^{-3}$ for the three groups, respectively. This aligns with our expectations, as recruiting more stable individuals is a slower process than recruiting less stable ones.

\begin{figure}[ht]
    \centering
    \begin{subfigure}[b]{.49\textwidth}
    \includegraphics[width=0.9\textwidth]{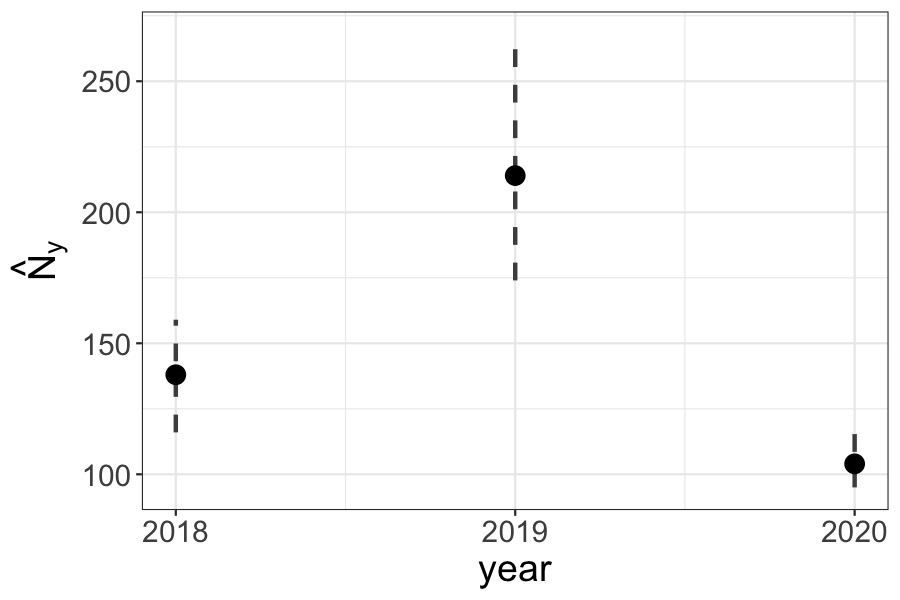}
    \caption{}
    \label{fig:Nocc_RealData}
    \end{subfigure}
    \begin{subfigure}[b]{.49\textwidth}
    \includegraphics[width=0.95\textwidth]{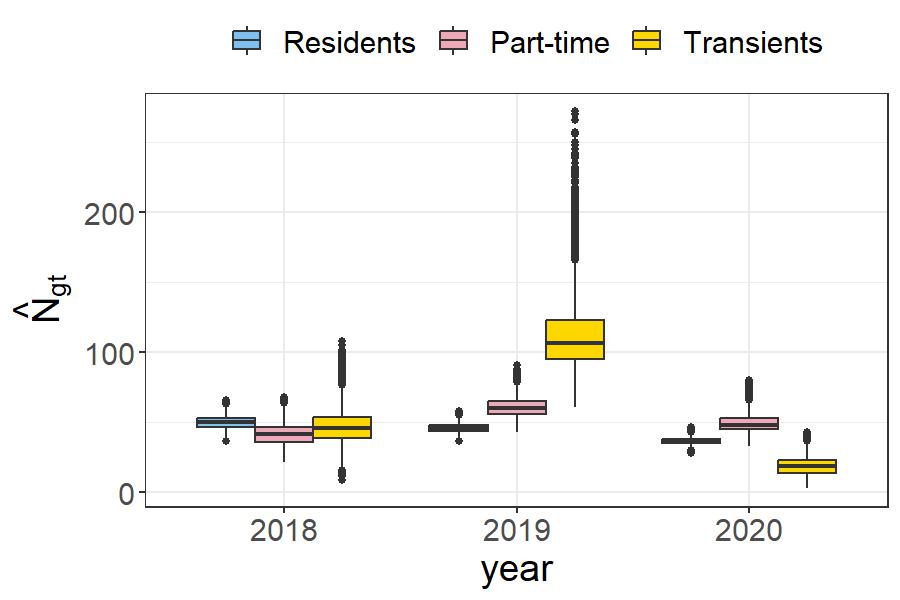}
    \caption{}
    \label{fig:Ngtocc_RealData}
    \end{subfigure}
    \caption{Point estimates and posterior $95\%$ credible intervals of the yearly super-population size (a) and by group (b).}
    \label{fig:Nyear_RealData}
\end{figure}

The final abundance estimate of the super-population in the whole observation window is of $\hat{N}_{super}=311$ ($CI_{0.95}=[266, 373]$), with yearly variations $\hat{N}_y$ ($y=2018,2019,2020$) that show a peak in 2019 and a decrease in the last year of observation (see Figure \ref{fig:Nocc_RealData}). It is, however, interesting to look at Figure \ref{fig:Ngtocc_RealData}, where we report the yearly abundance estimates by group to better understand the behaviour of the aggregated yearly pattern in Figure \ref{fig:Nocc_RealData}). 
Indeed, it seems that transients' abundance ($\hat{N}_{T}=174$ ($CI_{0.95}=[123, 236]$) is the main factor affecting the overall yearly counts, especially in 2019 and 2020. They show a substantial decrease in 2020, while the abundances of residents ($\hat{N}_{R}=57$ ($CI_{0.95}=[48, 65]$) and part-times ($\hat{N}_{P}=81$ ($CI_{0.95}=[55, 101]$) group tend to remain stable across years. 
Recall that these estimates only refer to well-marked individuals, who represent only a proportion of the whole population ($\approx 60\%$ of all sighted individuals across the three years)\footnote{Inferring the overall size from the well-marked abundance estimate is not the focus of this paper. For details on this passage, see \cite{wilson1999estimating}}. 
The posterior distributions of the yearly sizes of the three groups were obtained by counting all the individuals present in the population at each iteration by year and group.

By assigning the $195$ well-marked individuals observed between June 2018 and November 2020 to a single group according to the \textit{Maximum a Posteriori} (MAP) allocation, we have that $51$ are assigned to the group of residents, $54$ to the group of part-time and $90$ to the group of transients. 
Notice that the MAP is a straightforward and well-established method to attain classification in finite mixture modelling. However, it is known to be sub-optimal in some contexts \citep{stephens2000dealing,mclachlan2019finite} and other appealing methods have been proposed in the recent literature (e.g. \cite{wade2018bayesian}). The problem of summarising membership probabilities into a crisp classification is a challenging and long-debated issue on which there is no general agreement. Nevertheless, results from the simulation study (see Section \ref{sec:simstudy}) showed that the MAP is sufficiently reliable in the proposed framework. On the other hand, we are aware that any procedure deriving a crisp classification from membership probabilities over-simplifies the complexity of the inferred results as it does not give any information about the strength each individual is assigned to a specific group. 
The latter is one of the main advantages of the soft clustering returned by finite mixture models, and we wish to exploit its fuzziness to quantify how decisively each individual is assigned to one group or the other. 
For instance, $90\%$ of the individuals classified into the Resident group have been assigned to it with a probability greater than $0.9$. This probability is greater than $0.54$ for the Part-Times and $0.53$ for the Transient. 

\begin{figure}[ht]
    \centering
    \includegraphics[width=0.98\textwidth]{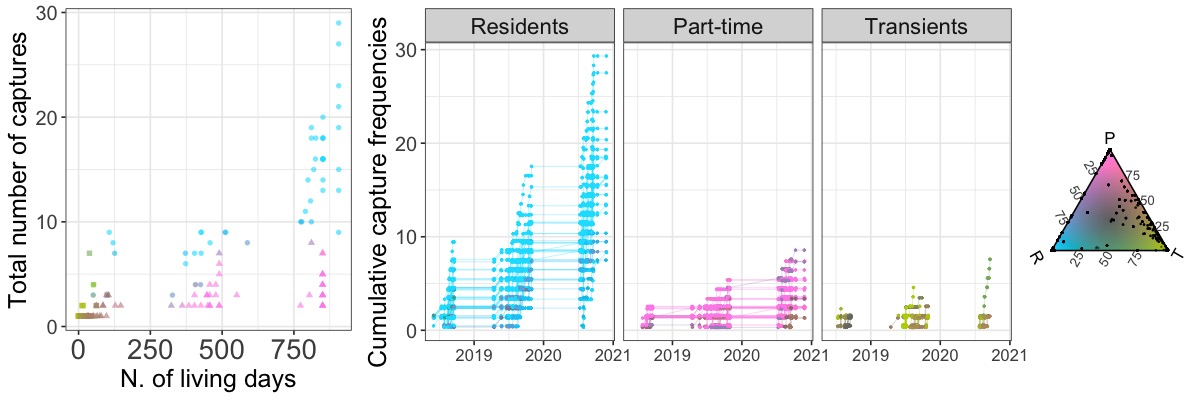}
    \caption{Individual cumulative frequencies of capture for all the encountered individuals divided into the three groups defined by the RPT model. Posterior allocation was based on MAP.}
    \label{fig:postalloc}
\end{figure}

In this regard, it is possible to visualise the classification results in Figure \ref{fig:postalloc} using a \textit{ternary diagram}, which allows quantifying the probability that each individual belongs to each group. The figure highlights three well-distinguishable capture history patterns that again comply with the typical RPT behaviour. 
Residents are available in the area for the whole study period and are spotted very frequently in subsequent sampling occasions; their capture histories are very informative. Hence, their identification is clear-cut (intense colour) in most cases. 
Part-time individuals are available in the area for most of the study period and are frequently spotted, but not as often as residents. Their classification is crisp if they have been encountered for the first time toward the beginning of the study period, while it is dicey when they have been encountered toward the end for the first time. Finally, the transients show short capture histories with few captures that never cross two years; this reflects that they are spotted only a few times and do not visit (\textit{survive}) the area for long. Given the little information provided by their capture histories, their classification is slightly vaguer compared to the other groups, especially for individuals only observed in the last sampling occasions.

Finally, it is important to remark that the proportions of the three groups do not align with the estimated weights $\hat{w}=(0.09, 0.39, 0.52)$. This is because the prior weights are not a good indicator of the true impact of each component in the super-population, but only of their impact on the augmented dataset (as mentioned in Section \ref{sec:simstudy}). 
However, we can infer on the composition of the $N-D$ uncaught individuals, namely those individuals that were recruited (i.e. not pseudo-individuals) but were never captured. The model allocates (on average) the uncaught individuals to the resident group for the $2\%$, the part-time group for the $15\%$ and the transient group for the $83\%$. This result complies with the short-term survival of the transient individuals and the more elusive nature of the part-time ones. On the other hand, the residents are always present in the area and, therefore, more susceptible to capture in a broad sense. The results suggest that most of the resident individuals ($98\%$) have already been observed and documented, a result which is in line with the progressive decrease of the discovery rate highlighted in Section \ref{sec:Motivating} (cfr. Figure \ref{cumdisc}).

\section{Discussion} \label{sec:discussion}

Estimating the abundance of marine wildlife species is a challenging but critical activity that can tell much about the undergoing ecological processes. Thus, combining high-quality data with solid analytical approaches is essential to improve our knowledge of these dynamics and increase the potential for management actions \citep{vella2021conservation, 10.3389/fmars.2022.782680}.
In this work, we pursued a Bayesian estimation of the size of a common bottlenose dolphin population, which is organised into three groups with different residency and site-fidelity patterns.
Accounting for such unobserved heterogeneity is a very common problem in the environmental literature. However, few papers approach the problem from the Bayesian perspective and develop ad-hoc solutions based on prior scientific knowledge of the population of interest. 

We proposed a parsimonious specification of a finite mixture model within the PX-DA setting for CR analysis, which we named RPT. This specification reflects the typical bottlenose dolphin residency pattern, with individuals showing high, partial or low site-fidelity \citep{dinis2016bottlenose, hunt2017demographic, haughey2020photographic, la2022determinants}. Its adoption, characterised by fewer free parameters, simplifies the identification of all the model components compared to more generic and flexible alternatives. Furthermore, while the application of finite mixture models to bring evidence about the true existence of different population groups has sometimes been discouraged \citep{pledger2000unified,pledger2003open}, here we have shown how they can be exploited to identify classes of individuals sharing similar profiles whenever strong scientific evidence of population groups' existence is available.

We devoted part of the model description to the discussion of a suitable prior elicitation preventing the label-switching issue of FMM. Formal derivations have enriched the results on constrained Beta priors from \cite{alaimodiloro2022}. The performances of our proposal have been assessed through a simulation study that considered scenarios with the increasing length of the observation window and, hence, of the capture histories (i.e. mimicking continuous monitoring of a population of interest over time). When data were generated from the RPT model, we evaluated the influence of the number of sampling occasions on the estimates' quality. As expected, yielded better accuracy and coverage thanks to the larger sample size. Comparison with alternative model versions over many replicas showed that the model estimation exhibited satisfying and robust performances when the observation window was long enough (more than $1$ year of monitoring).

We estimated the RPT model and other alternatives on the data that motivated our work. The results showed that, in terms of the WAIC score, our model outplays several well-established competitors in the class of JS-type models. 
The results correspond to biologically meaningful findings and align with previous research work.
The estimated abundance is of $\hat{N}^*_{super}=311$, of which $51$ are estimated to be resident members of the population. This quantity is particularly important as it is a proxy of the breeding population in the area. The yearly trend shows an increase between 2018 and 2019 and a decrease between 2019 and 2020.
While the estimated overall abundances are close to the ones obtained by \cite{pace2021capitoline}, the trend is slightly in contrast with their results (non-significant increasing trend between 2019 and 2020).
This difference must be due to the better quantification of the overall uncertainty in our analysis. This is particularly important in this context as the major interest lies in estimating a reasonable range that accounts for both the \textit{worst} and \textit{best} case scenario. Indeed, we estimate that the driving factor of the variations in the population size across the years is due to the abundance variation of the transient individuals, the proportion of which is dynamic and not static. Changes in the presence of such dolphins in an area may reflect changes in driving features like habitat quality, prey distribution, or anthropogenic disturbance. The year 2020 was the COVID-19 pandemic, with changes in ecological conditions and human pressures on coastal marine waters \citep{carome2022long}. Although difficult to assess, the influence of these factors on the number of transient individuals cannot be ruled out, as several dolphins may have been induced to reduce their mobility toward the study area by improved conditions in their native habitat.

One basic assumption of CR experiments that we do not drop is that captures of different individuals are independent. However, it is well known that bottlenose dolphin populations can form structured societies with complex social networks \citep{pace2022resources}.  In future studies, we would like to drop the independence assumption and include information on the population's social structure and the consequent statistical dependence in the capture histories. Including the effect of external or individual covariates would also be interesting. For instance, it is possible to recognise the gender and age class in high-quality pictures. This partial information may be incorporated in the Bayesian framework and could help for a better assessment, for example, of the membership of different individuals in different groups or of their marking probability \citep{wu2021bayesian}. The capture-recapture model to estimate the abundance of the bottlenose dolphins could be extended by adopting a stopover model \citep{pledger2009stopover,worthington2019estimating}, which allows capture and survival probabilities to depend both on time and time since arrival in the population. In that case, the model could be formulated as a multi-state Hidden Markov model \citep{worthington2019estimating}, where the different states refer to the different groups of individuals in the population. Last but not least, it would be extremely interesting to conduct the survey on a larger spatial scale and include external information to account for the spatial heterogeneity \citep{wu2017bayesian}.



\section*{Conflict of interest statement}

The authors declare no conflict of interest.

\section*{Data availability statement}

The data that supports the findings of this study are available in the supplementary material of this article.


\appendix
\section*{Appendix}
\sectionfont{\normalsize}
\subsectionfont{\small}

\section{Multi-state interpretation of the ecological process}\label{multistate}

The ecological model described by \cite{royle2008hierarchical} can be also seen as a multi-state process. At each occasion $t$ each individual in the augmented matrix can be in one and only one of the following three states:
\begin{enumerate}
    \item it has never been part of the population
    \item it is part of the population
    \item it was part of the population, but now it is not
\end{enumerate}

We first notice that when an individual becomes part of the population, it cannot be recruited any more: following the notation introduced in Section \ref{subsec:JSintro}, this implies that for $t>1$, $r_{it}$ and $z_{it}$ cannot be simultaneously equal to $1$.

In the JS modelling framework individuals that leave the population cannot return to it. Hence, state $3$ is an absorbing state.
Let us momentarily ignore the population heterogeneity (clustering structure) for the sake of clarity. If we allow temporal heterogeneity, then the transition probability matrix associated with the three states at times $t=2,\dots,T$ is:
\begin{equation*}
    \begin{blockarray}{c@{\hspace{1pt}}ccc@{\hspace{3pt}}}
         & 1   & 2   & 3 \\ 
        \begin{block}{c@{\hspace{3pt}}
    |@{\hspace{2pt}}ccc@{\hspace{2pt}}|}
        1 & 1-\rho_t & \rho_t & 0 \\
        2 & 0 & \phi_t & 1-\phi_t \\
        3 & 0   & 0   & 1   \\
       \end{block} 
    \end{blockarray} 
\end{equation*}
where rows and columns represent the states at time $t$ and $t+1$, respectively. At period $t=1$, all individuals can be recruited in the population. As $t$ increases, more and more individuals enter the population or, equivalently, are \textit{removed} from state 1. All the observed individuals will eventually be recruited in the population before time $T$, but not all of them will leave it (may have survived to future, unobserved, periods). 
Remember that individuals are exposed to capture, with probability $p_t$, only during their transitory stay in state 2. Hence, also a portion of never observed individuals may have been recruited into the population at some point without ever being captured. They represent the unknown part we aim to estimate.

\section{More details on the prior specification}\label{app:betapriors}

The standard solution is that of concatenating conditionally specified Uniform distributions as follows:
\begin{equation}
\label{eq:unitPrior}
        u_1\sim Unif(0, 1),\quad u_j|u_{j-1}\sim Unif(u_{j-1}, 1)%
    \qquad j=2,\dots, G,
\end{equation}
where the $u_j$'s are, for instance, the survival probabilities \citep{turek2021bayesian}.
While effective in imposing the constraint, the priors in \eqref{eq:unitPrior} do not allow for the inclusion of previous information that can ease parameters' identification. \cite{alaimodiloro2022} explore alternative conditional prior specification that, while implementing the ordering constraint, allows to control for the shape and first moments of the induced marginal prior distributions:
 $$   \pi_{u_g}(u_g) = \int_{\mathcal{S}_{g-1}}\pi_{u_1}(u_1)\prod_{j=2}^{g}\pi_{u_j|u_{j-1}}(u_j\,|\,u_{j-1})\,du_1,\dots,u_{g-1}\,,$$
where $\mathcal{S}_{g-1}$ is a simplex of order $g-1$. 
Possible choices are the Beta and Truncated Beta distributions. The latter corresponds to the following set of prior distributions:
\begin{equation}
\label{eq:betatPrior}
    \begin{aligned}
    & u_1\sim Beta(\alpha_1,\,\beta_1), \, \quad 
    & u_g|u_{g-1}\sim tBeta(\alpha_g,\,\beta_g; u_{g-1},1)
    \end{aligned}\qquad g=2,\dots, G,
\end{equation}
where $tBeta(\alpha_g,\,\beta_g; l, 1)$ denotes the Truncated Beta distribution in $(l, 1)$. Note that for $\alpha_g = \beta_g=1$ we obtain \eqref{eq:unitPrior}.
When $G=2$ and $\alpha_2=1$, the prior specification in \eqref{eq:betatPrior} induces the following marginal prior density on $u_2$:
\begin{equation}
\label{eq:tbetaMarg}
    \pi_{u_2}(u_2)= \frac{B(\alpha_1,\beta_1-\beta_2)}{B(\alpha_1,\beta_1)}\,\beta_2(1-u_2)^{\beta_2-1}\,F_{Beta(\alpha_1,\beta_1-\beta_2)}(u_2)\,,\quad\text{with } \beta_1>\beta_2\,\,,
\end{equation}
where $F_{Beta(\alpha_1,\beta_1-\beta_2)}(\cdot)$ is the cdf of a $Beta(\alpha_1, \beta_1-\beta_2)$. See Section \ref{truncBeta} for the formal proof. 
When $\alpha_1=\beta_2=k$ and $\beta_1=k+1$, we have that $\pi_{u_1}(u_1)=Beta(u_1\,|\,k,\,k+1)$ and the density in  \eqref{eq:tbetaMarg} is a $Beta(k+1,\,k)$. The two distributions are mirrored with respect to the vertical line $v=0.5$ (see Figure \ref{fig:tBetaprior} of the Appendix as an example). 
Hence, we have $\mathbb{E}[u_2]=1-\mathbb{E}[u_1]$, for all $k>0$; if $k>1$ we also have $Mode(u_2)=1-Mode(u_1)$.
The low-parametrised structure induces well-separated prior means or prior modes for $u_1, u_2$ marginal distributions, favoring the mixture components' separation.
Alternative settings inducing well separated modes in beta-type priors are reported in Section \ref{restrBeta} of the Appendix.

\subsection{The Beta and Truncated Beta distribution}\label{truncBeta}

The density of a Beta random variable with shape $\alpha$ and rate $\beta$ is:
\begin{equation*}
    Beta\lrnd z\given \alpha, \beta\rrnd = \frac{\Gamma\lrnd\alpha+\beta\rrnd}{\Gamma(\alpha)\Gamma(\beta)}\cdot z^{\alpha-1}\cdot (1-z)^{\beta-1}, \quad z\in(0, 1),
\end{equation*}
where $\Gamma(\cdot)$ is the Euler Gamma function.
The truncation requires normalising the same distribution over the truncated domain.
We say $u_2|u_1$ is a truncated Beta in $(u_1,1)$ when:
\begin{equation*}
    \pi(u_2|u_1)=tBeta(\alpha_2,\beta_2,\, u_1, 1)=\frac{1}{B(\alpha_2,\beta_2)}\,\frac{u_2^{\alpha_2-1}\,(1-u_2)^{\beta_2-1}}{1-F_{Beta(\alpha_2,\beta_2)}(u_1)}\,\mathbb{1}_{\{u_1,1\}}(u_2)\,,
\end{equation*}
with $F_{Beta(\alpha_2,\beta_2)}(\cdot)$ being the cdf of a $Beta(\alpha_2,\beta_2)$. 
Now, suppose that $u_1\sim Beta(\alpha_1,\beta_1)$ and $u_2|u_1\sim tBeta(\alpha_2,\beta_2,\, u_1, 1)$, then if $\alpha_2=1$ the marginal prior distribution induced on $u_2$ is given by:
\begin{align*}
    \pi(u_2) & = \int_0^1\,\pi(u_2|u_1)\pi(u_1)\,du_1  \\
    & = \int_0^1\,\frac{\beta_2\,(1-u_2)^{\beta_2-1}}{1-F_{Beta(1,\beta_2)}(u_1)}\,\mathbb{1}_{\{u_1,1\}}(u_2)\frac{1}{B(\alpha_1,\beta_1)}\times u_1^{\alpha_1-1}(1-u_1)^{\beta_1-1}\,du_1\\
    & = \frac{\beta_2\,(1-u_2)^{\beta_2-1}}{B(\alpha_1,\beta_1)}\,\int_0^{u_2}\,\frac{u_1^{\alpha_1-1}(1-u_1)^{\beta_1-1}}{1-F_{Beta(1,\beta_2)}(u_1)}\,du_1\\
    & = \frac{\beta_2\,(1-u_2)^{\beta_2-1}}{B(\alpha_1,\beta_1)}\,\int_0^{u_2}\,u_1^{\alpha_1-1}(1-u_1)^{(\beta_1-\beta_2)-1}\, d u_1\\
    & = \frac{B(\alpha_1,\beta_1-\beta_2)}{B(\alpha_1,\beta_1)}\,\beta_2\,(1-u_2)^{\beta_2-1}\,\int_0^{u_2}\,\frac{u_1^{\alpha_1-1}(1-u_1)^{(\beta_1-\beta_2)-1}}{B(\alpha_1,\beta_1-\beta_2)}\,du_1\\
    & = \frac{B(\alpha_1,\beta_1-\beta_2)}{B(\alpha_1,\beta_1)}\,\beta_2(1-u_2)^{\beta_2-1}\,F_{Beta(\alpha_1,\beta_1-\beta_2)}(u_2)\,,\quad {with }\,\, \beta_1>\beta_2,
\end{align*}
where we exploited the fact that $F_{Beta(1,b)}(z)=\int_0^z\,(1-t)^{b-1}\,dt=1-(1-z)^b$. Observe that the constraint $\beta_1>\beta_2$ is essential to avoid the divergence of the beta function.

In particular, it is straightforward to show that if $\alpha_1=\beta_2=k$ and $\beta_1=k+1$, that is $u_1\sim Beta(k,k+1)$ and $u_2|u_1\sim tBeta(1,k,\, u_1, 1)$, then $u_2\sim Beta(k+1,k)$.
Notice that, in this particular case, the marginal distribution induced on $u_2$ is symmetrical with respect to the distribution of $u_1$ around the vertical line $v=0.5$; equivalently, $u_2\overset{d}{=}1-u_1$.

\subsection{The Beta and Restricted Beta}\label{restrBeta} 
\begin{figure}[ht]
    \centering
    \begin{subfigure}[b]{.49\textwidth}
    \includegraphics[width = .9\textwidth]{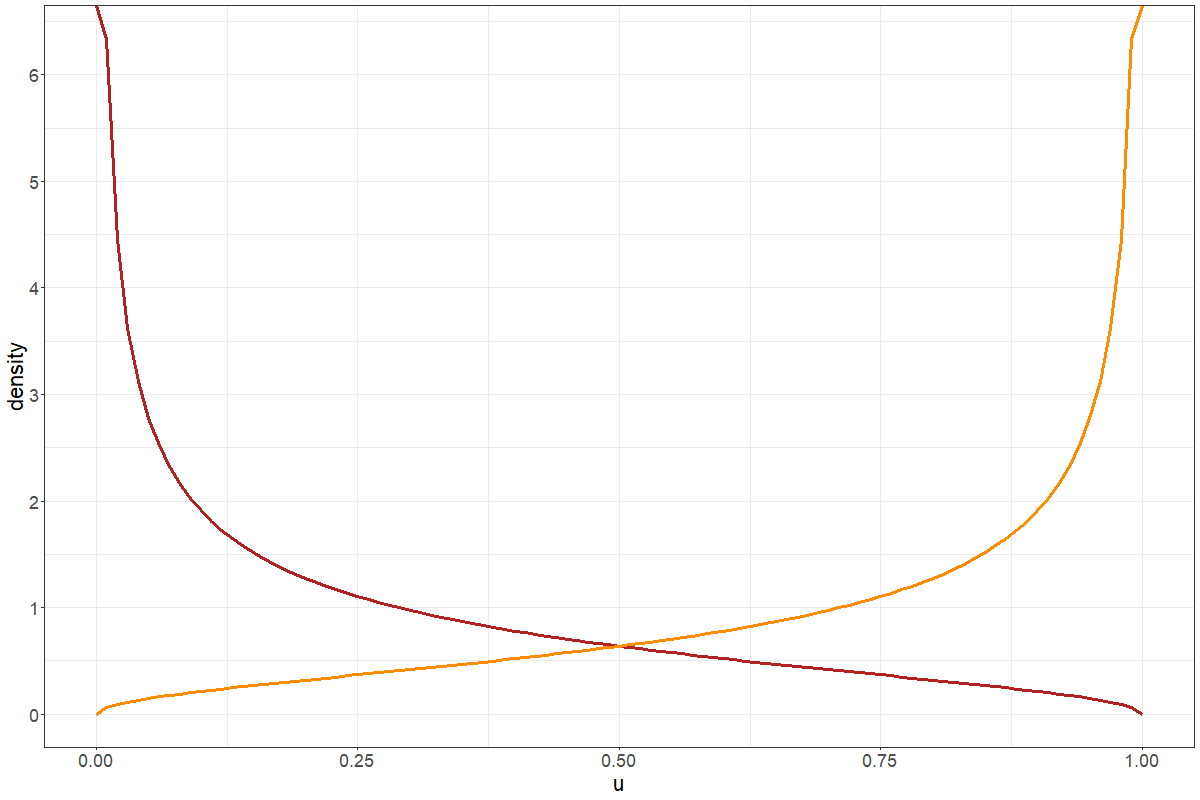}
    \caption{}
     \label{fig:tBetaprior}
    \end{subfigure}
     \begin{subfigure}[b]{.49\textwidth}
     \includegraphics[width = .9\textwidth]{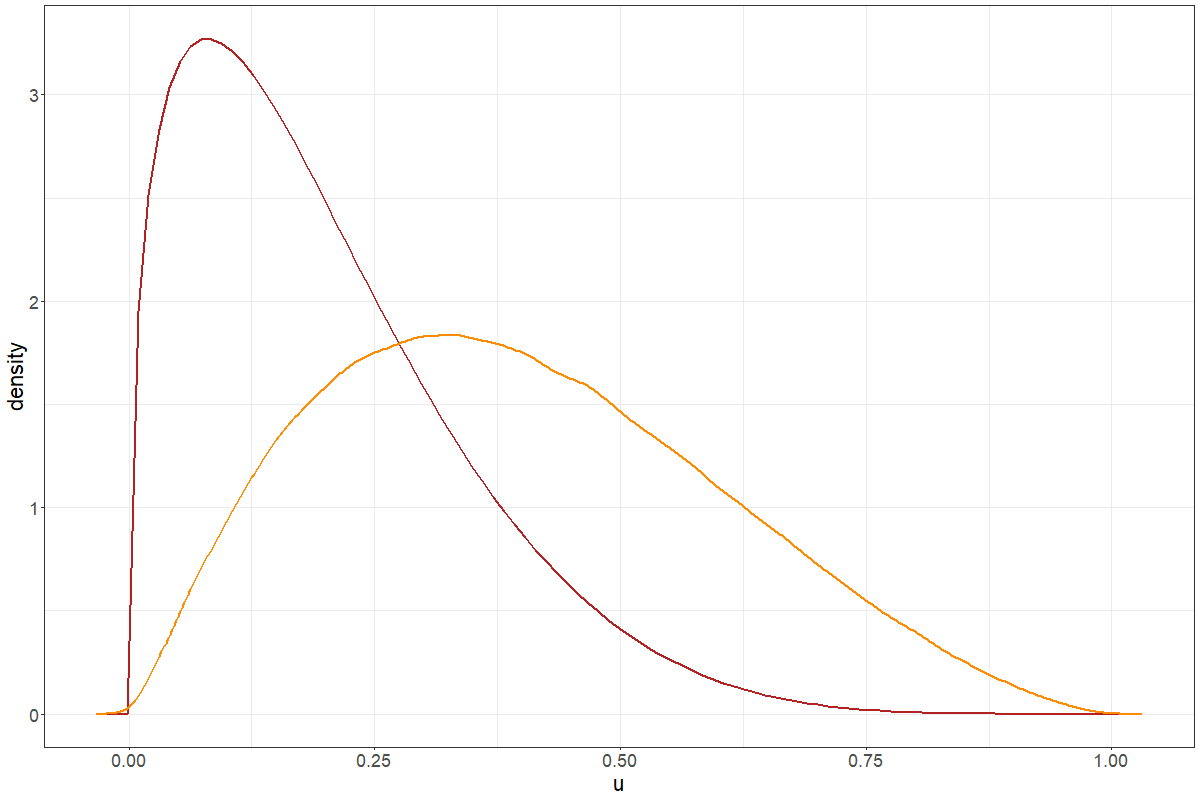}
     \caption{}
     \label{fig:rBetaprior}
    \end{subfigure}
    \caption{(a) Marginal priors induced on $u_1$ (solid red) and $u_2$ (dashed orange) by the conditional specification $u_1 \sim Beta(.5, 1.5),\; u_2|u_1 \sim tBeta(1, .5, u_1, 1)$. (b) Marginal priors induced on $u_1$ (solid red) and $u_2$ (dashed orange) with $\mathbb{E}[u_1]=0.2$, $\mathbb{E}[u_2]=0.4$ with $\mathbb{V}[u_1]=0.02$ and $\mathbb{V}[u_2]=0.04$ by the conditional prior specification $u_1\sim Beta(1.4, 5.6),\;u_2|u_1\sim rBeta(0.826, 2.478, u_1, 1)$ .}
    \label{fig:AllBetaprior}
\end{figure}

An alternative conditional specification that allows for a properly informed marginal prior can be obtained using Restricted Beta distribution. Also known as \textit{4-parameters Beta}, it is a Beta r.v. that has been shifted and scaled to reside on the domain $(l, u)$:
\begin{equation*}
    rBeta\lrnd z\given \alpha, \beta, l, u\rrnd = \frac{\Gamma\lrnd\alpha+\beta\rrnd}{\Gamma(\alpha)\Gamma(\beta)}\cdot \frac{(z-l)^{\alpha-1}\cdot (u-z)^{\beta-1}}{(u-l)^{\alpha+\beta-1}}\,\mathbb{1}_{\{l,u\}}(z).
\end{equation*}
We can use it to specify recursively a set of $G$ conditional priors as:
\begin{equation*}
\label{eq:betaRPrior}
    \begin{aligned}
    & u_1\sim Beta(\alpha_1,\,\beta_1), \quad 
    & u_g|u_{g-1}\sim rBeta(\alpha_g,\,\beta_g; u_{g-1},1)
    \end{aligned}\qquad g=2,\dots, G,
\end{equation*}
where $rBeta(\alpha_g,\,\beta_g; u_{g-1},1)$ denotes the Beta restricted to $(u_{g-1}, 1)$.
The corresponding joint prior is:
\begin{equation*}
    \pi\lrnd u_1, u_2\rrnd = \frac{\Gamma(\alpha_1+\beta_1)\Gamma(\alpha_2+\beta_2)}{\Gamma(\alpha_1)\Gamma(\beta_1)\Gamma(\alpha_2)\Gamma(\beta_2)}\cdot u_1^{\alpha_1-1}(1-u_1)^{\beta_1-\beta_2-\alpha_2}(u_2-u_1)^{\alpha_2-1}(1-u_2)^{\beta_2-1}\,\mathbb{1}_{\{u_1,1\}}(u_2),
\end{equation*}
which does not allow for an analytical marginalisation to get $\pi(u_2)$ in the general scenario. 
However, the $rBeta$ expected value and variance are available in closed form and hence we can use the \textit{law of total expectation} to derive the marginal expected value and variance of all the components. For $g=2,\dots,G$:
\begin{equation*}
\label{eq:resBetaMom}
\begin{aligned}
    &\bbE\lsq u_g\rsq = \frac{\alpha_g}{\alpha_g+\beta_g}+\mu_{g-1}\frac{\beta_g}{\alpha_g+\beta_g}\\
    &\bbV\lsq u_g\rsq = \sigma^2_{g-1}\cdot\frac{\beta_g^2}{(\alpha_g+\beta_g)^2}\lrnd 1+\frac{\alpha_g}{\beta_g(\alpha_g+\beta_g+1)}\rrnd+\lrnd 1-\mu_{g-1}\rrnd^2\cdot\frac{\alpha_g\beta_g}{(\alpha_g+\beta_g)^2(\alpha_g+\beta_g+1)},
\end{aligned}
\end{equation*}
where $\mu_{g-1}=\mathbb{E}[u_{g-1}]$ and $\sigma^2_{g-1}=\mathbb{V}[u_{g-1}]$.
Therefore, one can define a system of equations to find the combination of $\alpha_g, \beta_g$ that complies with a prior knowledge on the moments of the parameters (see Figure \ref{fig:rBetaprior} as an example).

\paragraph{A convenient parameter setting} A convenient parameter setting arises when $G=2$ and if we fix $\alpha_1=\alpha$, $\alpha_2=1$,  $\beta_1=\beta+1$ and $\beta_2=\beta$.
Indeed, if $u_1\sim Beta(\alpha, \beta+1)$ and $u_2\given u_1 \sim rBeta\lrnd 1, \beta, u_1, 1\rrnd$, then the joint distribution of $(u_1,u_2)$ is:
\begin{equation*}
    \pi\lrnd u_1, u_1\rrnd = \frac{(\alpha+\beta)\Gamma(\alpha+\beta)}{\Gamma(\alpha)\Gamma(\beta)}\cdot u_1^{\alpha-1}(1-u_2)^{\beta}.
\end{equation*}
Marginalising with respect to $u_1$ we get:
\begin{equation*}
    \begin{aligned}
        \pi\lrnd u_2\rrnd &= \int_0^1\pi(u_1, u_2)\,\mathbb{1}_{\{u_1,1\}}(u_2)\,d u_1=\int_0^1 \pi(u_1, u_2)\,\mathbb{1}_{\{0,u_2\}}(u_1)\,d u_1=\int_0^{u_2} \pi(u_1, u_2)\, d u_1=\\
        &= \frac{(\alpha+\beta)\Gamma(\alpha+\beta)}{\Gamma(\alpha)\Gamma(\beta)}\cdot (1-u_2)^{\beta-1}\cdot \int_0^{u_2} u_1^{\alpha-1}\,d u_1=\\
        &= \frac{\Gamma(\alpha+\beta+1)}{\Gamma(\alpha)\Gamma(\beta)}\cdot (1-u_2)^{\beta-1}\cdot \frac{u_2^{\alpha}}{\alpha}=\\
        &= \frac{\Gamma(\alpha+\beta+1)}{\Gamma(\alpha+1)\Gamma(\beta)}\cdot u_2^{(\alpha+1)-1}(1-u_2)^{\beta-1},
    \end{aligned}
\end{equation*}
which is a standard Beta density $\pi\lrnd u_2\rrnd = Beta(u_2\given \alpha+1, \beta)$. Expected value and variance can then be derived from basic properties of the Beta distribution.

\section{Modelling the temporary emigration}\label{app:tempem}

Here, we show that the introduction of the \textit{undetectability parameter} $\delta$ on the part-time individuals is equivalent to allowing for random temporary emigration.
We focus on the higher hierarchy level of the part-time model specification, as all other components are not affected. For the sake of clarity, we drop the $g$-subscript and let the reference to the part-time group be implied.

The proposed model specification for the part-time group is as follows:
\begin{equation}
\label{eq:undp}
\begin{aligned}
    &y_{it}\,|\, z_{it}\,\sim\, Bern(p_{t}\cdot (1-\delta)\cdot z_{it}),\\
    &z_{it}\,|\, z_{i,t-1},r_{it}\,\sim\, Bern(\phi_{t}\cdot z_{i,t-1}+\rho_{t}\cdot r_{it}), \quad  r_{it}=\text{min}\{r_{i,t-1},1-z_{i,t-1}\},
\end{aligned}
\end{equation}
which is different from the other groups only through the introduction of the parameter $\delta\in(0, 1)$ in the detection process.

Let us recall that $z_{it}$ is a latent variable indicating whether individual $i$ is \textit{``alive''} at time $t$. This is generally confounded with permanent emigration, but it cannot account for temporary emigration as exited individuals cannot ever re-enter the study area and return susceptible to captures. Therefore, this first latent variable is only able to model the time at which individual $i$ starts visiting the area (is born) and the time at which it stops visiting it for good (dies). 
The explicit modelling of temporary emigration within this time window requires the introduction of an additional latent variable $v_{it}$ denoting whether individual $i$ is present given that $z_{it}=1$. 
The specification of Equation \eqref{eq:undp} arises if we assume that temporary emigration occurs randomly and with equal probability $\delta$ while individual $i$ is alive. This corresponds to the following hierarchical specification:
\begin{equation}
\label{eq:undfull}
\begin{aligned}
    &y_{it}\,|\, v_{it}\,\sim\, Bern(p_{t}\cdot v_{it}),\\
    &v_{it}\,|\, z_{i,t}\,\sim\, Bern((1-\delta)\cdot z_{i,t}),\\
    &z_{it}\,|\, z_{i,t-1},r_{it}\,\sim\, Bern(\phi_{t}\cdot z_{i,t-1}+\rho_{t}\cdot r_{it}), \quad  r_{it}=\text{min}\{r_{i,t-1},1-z_{i,t-1}\},
\end{aligned}
\end{equation}
where $y_{it}\perp z_{it}$ if $v_{it}$ is known.
We can easily marginalise $v_{it}$ out of Equation \eqref{eq:undfull} by noting that each $y_{it}|z_{it}$ is a Bernoulli random variable with probability of success:
\begin{equation*}
\begin{aligned}
\tilde{p}_t&=\bbE\lsq y_{it}\,|\,z_{it}\rsq\,=\,\bbE\lsq\bbE\lsq y_{it}\,|\,v_{it}\rsq\,|\,z_{it}\rsq=\\
&=\bbE\lsq p_t\cdot v_{it}\,|\,z_{it}\rsq=p_t\cdot\bbE\lsq v_{it}\,|\,z_{it}\rsq=p_t\cdot (1-\delta)\cdot z_{it},
\end{aligned}    
\end{equation*}
from which:
$$
y_{it}\,|\, z_{it}\,\sim\, Bern(p_{t}\cdot (1-\delta)\cdot z_{it})
$$

This equivalence and the corresponding interpretation are what motivates the use of a multiplicative parametrisation on the capture probability at the invlogit scale in place of a more straightforward group-specific intercept within the logit specification.

\section{Details about time lags used in the simulation study}\label{app:timelags}

The number of days between two consecutive capture occasions (\textit{daily time lags}, henceforth) within a single year has been simulated from a \textit{shifted} geometric distribution with probability $0.05$, which has an expected value equal to $20$ and a standard deviation equal to $19.5$. The resulting random sequence of time lags is $$(20,1,12,15,56,9,9,12,10)$$ and, for scenarios that contemplate more than one year of study, the same sequence is repeated during each new year. The shift of year occurring each $10$ occasions is achieved by using a higher constant time lag (i.e. $240$ days) between the $(10k)$th occasion and the $(10k+1)$th occasion, with $k=1,2,3$. This results in a scenario $k$ composed by $k$ years of study, for $k=1,2,3,4$. For example, scenario 2 ($T=20$) is composed of the following sequence of time lags, resulting in $2$ years of capture occasions: $$(20,1,12,15,56,9,9,12,10,240,20,1,12,15,56,9,9,12,10)$$

\section{Convergence of relevant parameters estimated on real data}\label{app:diagnostics}

We checked the convergence of the parameter chains in the real data application through the general-purpose Gelman diagnostic.
All potential scale reduction factors $\hat R$ are below $1.01$, which suggests a good mixing of all parameter chains. 

We show the behaviour of the traceplots and density plots of the most relevant time-static parameters for the sake of saving space.
We can observe how the two chains explore the same parameters space in all cases and produce a well-shaped posterior distribution, with no bad behaviour.

\begin{figure}[ht]
    \centering
    \begin{subfigure}[b]{.24\textwidth}
    \includegraphics[width=0.9\textwidth]{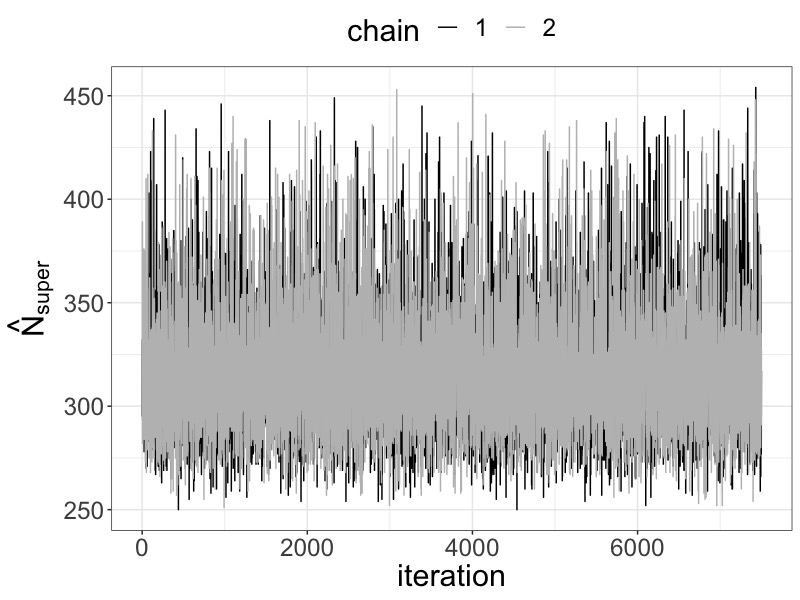}
    \caption{}
    \label{fig:Nsup_trace}
    \end{subfigure}
    \begin{subfigure}[b]{.24\textwidth}
    \includegraphics[width=0.9\textwidth]{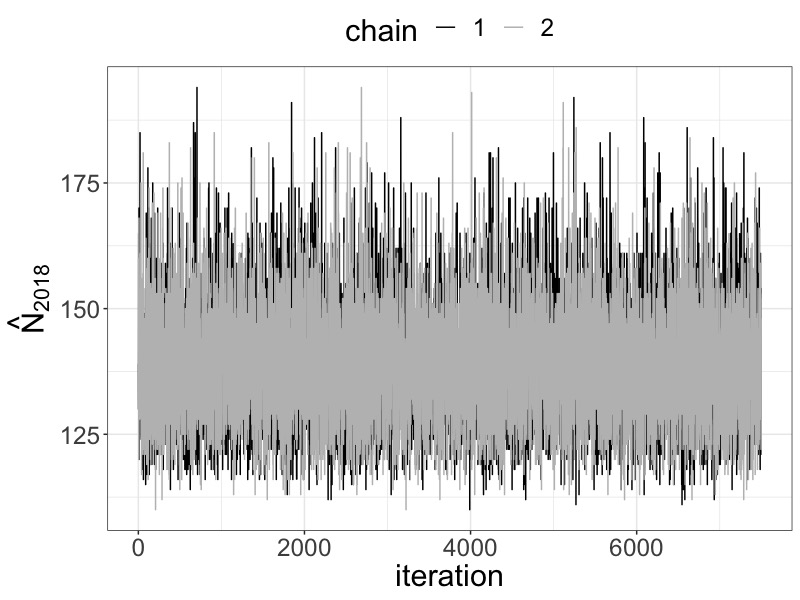}
    \caption{}
    \label{fig:N1_trace}
    \end{subfigure}
    \begin{subfigure}[b]{.24\textwidth}
    \includegraphics[width=0.9\textwidth]{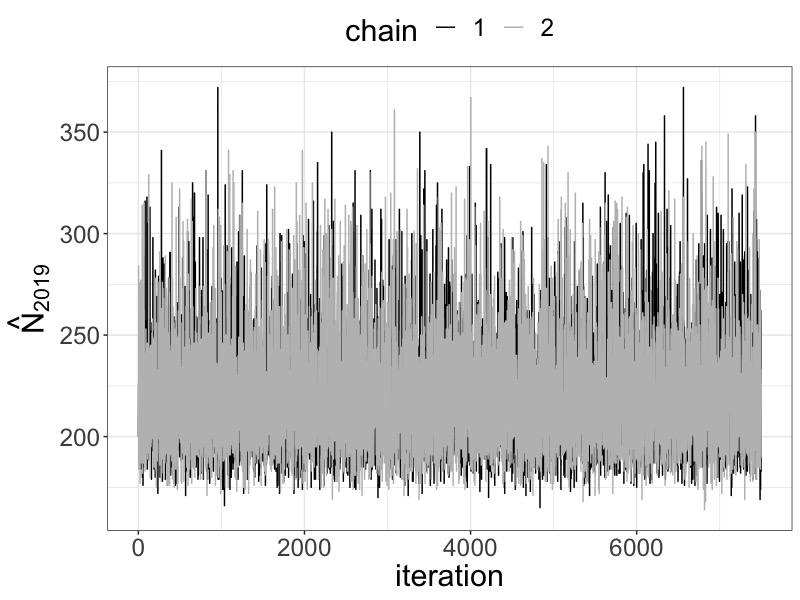}
    \caption{}
    \label{fig:N2_trace}
    \end{subfigure}
    \begin{subfigure}[b]{.24\textwidth}
    \includegraphics[width=0.9\textwidth]{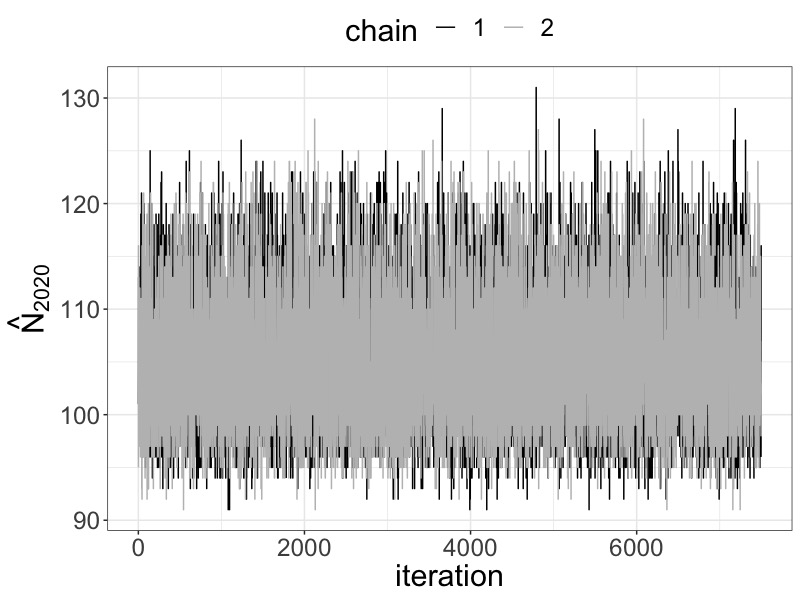}
    \caption{}
    \label{fig:N3_trace}
    \end{subfigure}
    \begin{subfigure}[b]{.24\textwidth}
    \includegraphics[width=0.9\textwidth]{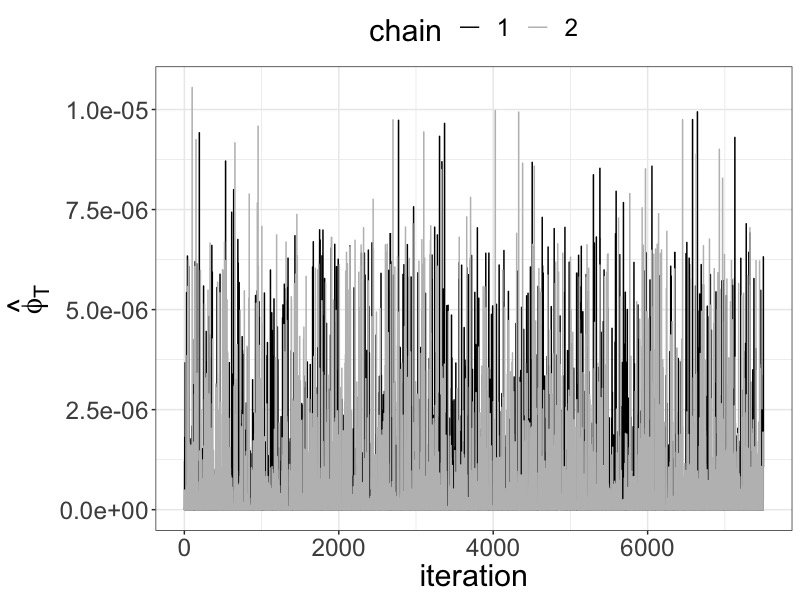}
    \caption{}
    \label{fig:phi1_trace}
    \end{subfigure}
    \begin{subfigure}[b]{.24\textwidth}
    \includegraphics[width=0.9\textwidth]{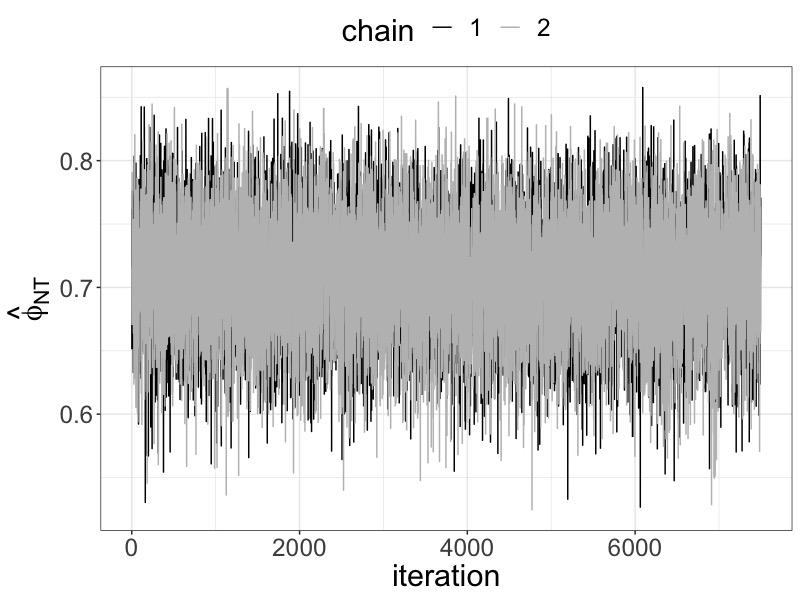}
    \caption{}
    \label{fig:phi2_trace}
    \end{subfigure}
    \begin{subfigure}[b]{.24\textwidth}
    \includegraphics[width=0.9\textwidth]{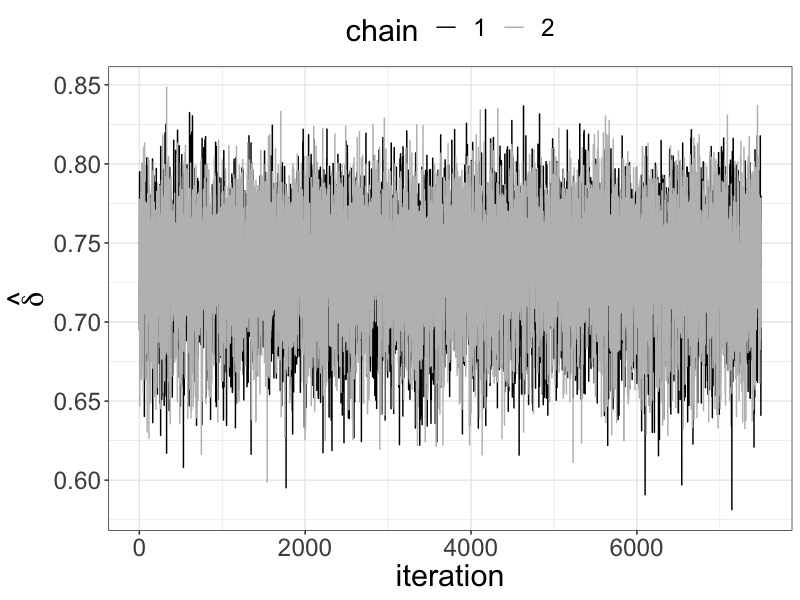}
    \caption{}
    \label{fig:delta_trace}
    \end{subfigure}
    \begin{subfigure}[b]{.24\textwidth}
    \includegraphics[width=0.9\textwidth]{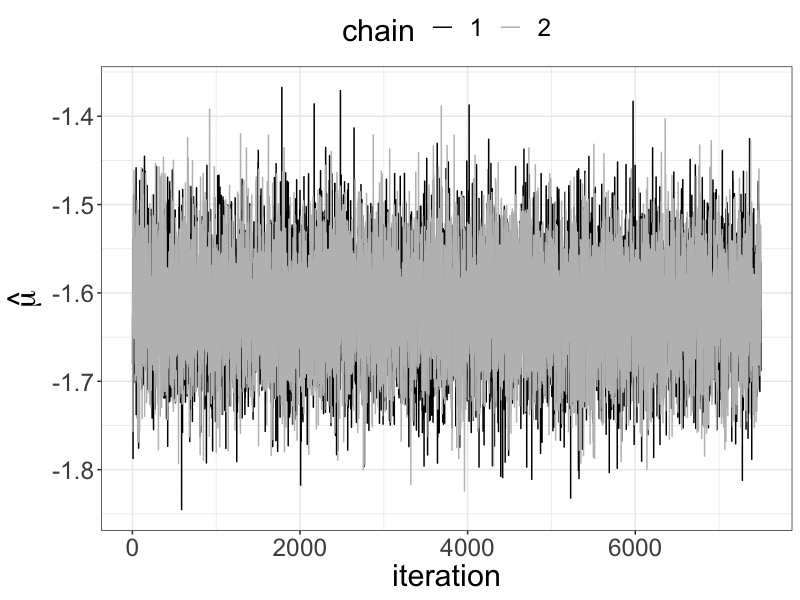}
    \caption{}
    \label{fig:mu_trace}
    \end{subfigure}
    \caption{Traceplots of posterior samples for the main parameters of interest of model RPT.}
    \label{fig:traceplots}
\end{figure}

\begin{figure}[ht]
    \centering
    \begin{subfigure}[b]{.24\textwidth}
    \includegraphics[width=0.9\textwidth]{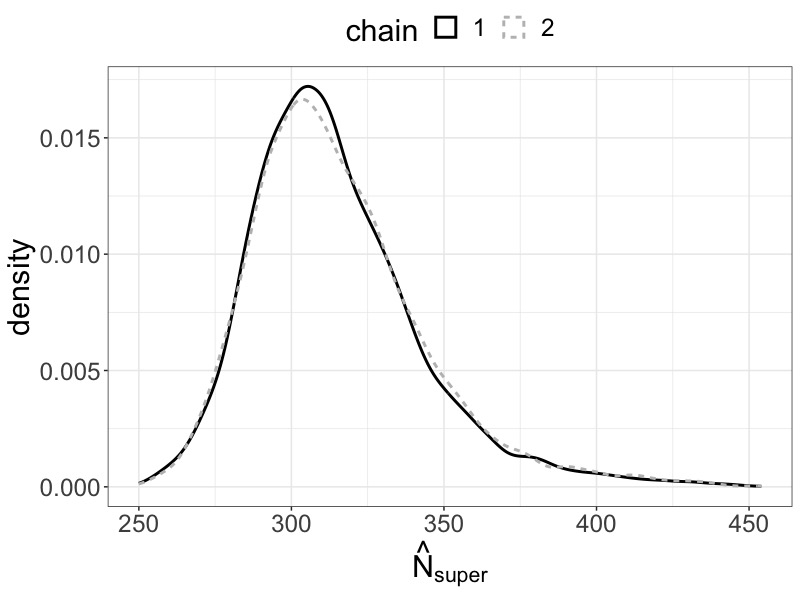}
    \caption{}
    \label{fig:Nsup_dens}
    \end{subfigure}
    \begin{subfigure}[b]{.24\textwidth}
    \includegraphics[width=0.9\textwidth]{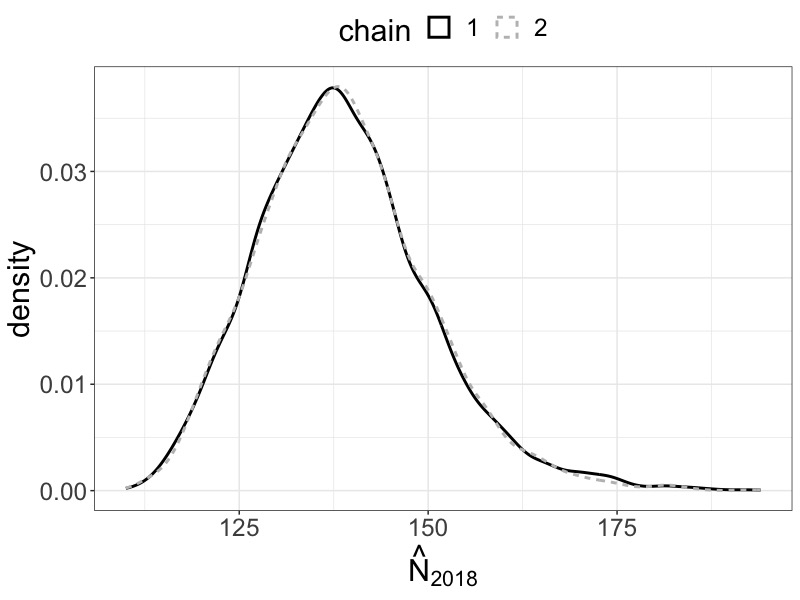}
    \caption{}
    \label{fig:N1_dens}
    \end{subfigure}
    \begin{subfigure}[b]{.24\textwidth}
    \includegraphics[width=0.9\textwidth]{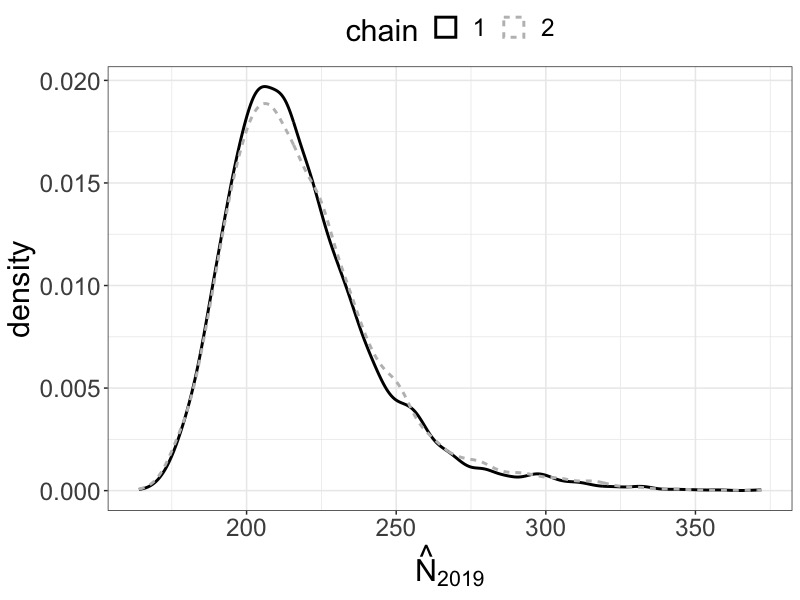}
    \caption{}
    \label{fig:N2_dens}
    \end{subfigure}
    \begin{subfigure}[b]{.24\textwidth}
    \includegraphics[width=0.9\textwidth]{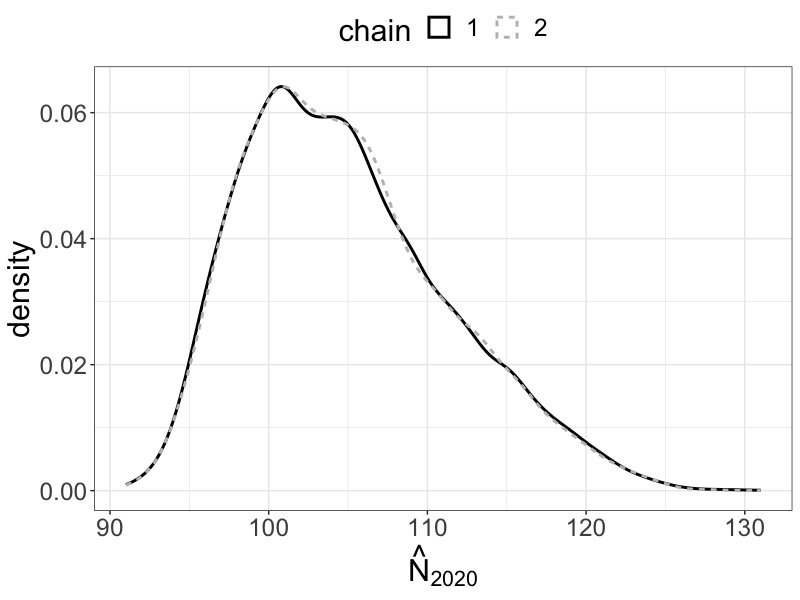}
    \caption{}
    \label{fig:N3_dens}
    \end{subfigure}
    \begin{subfigure}[b]{.24\textwidth}
    \includegraphics[width=0.9\textwidth]{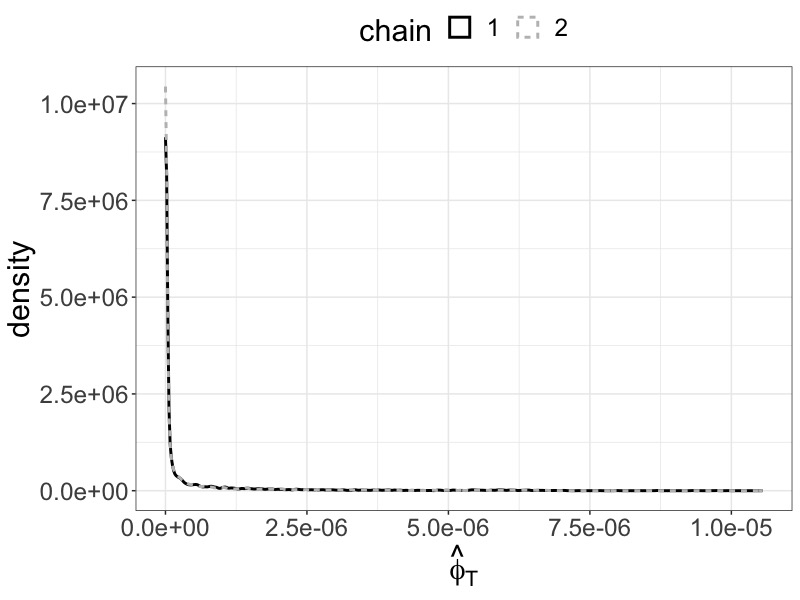}
    \caption{}
    \label{fig:phi1_dens}
    \end{subfigure}
    \begin{subfigure}[b]{.24\textwidth}
    \includegraphics[width=0.9\textwidth]{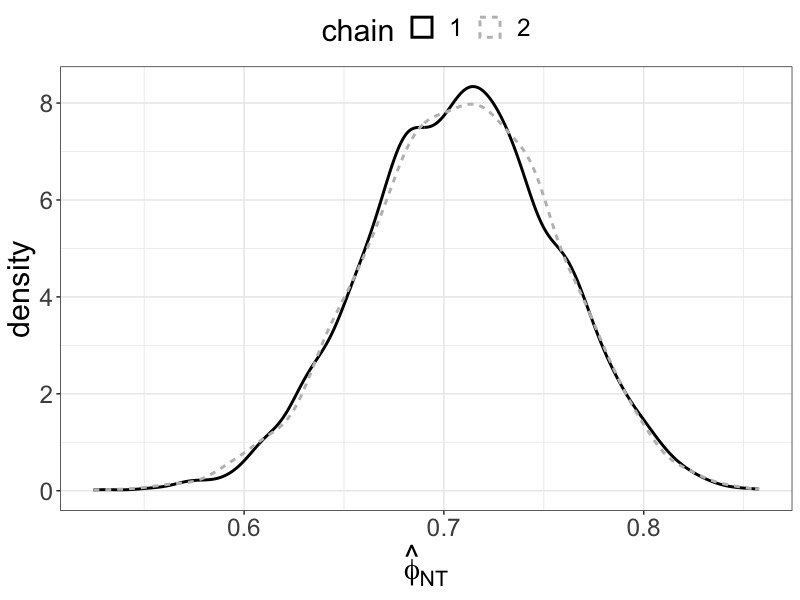}
    \caption{}
    \label{fig:phi2_dens}
    \end{subfigure}
    \begin{subfigure}[b]{.24\textwidth}
    \includegraphics[width=0.9\textwidth]{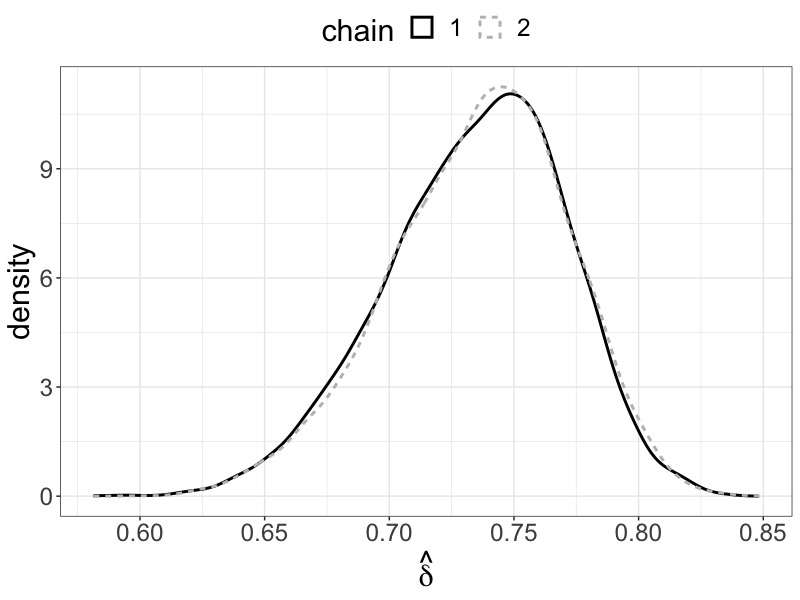}
    \caption{}
    \label{fig:delta_dens}
    \end{subfigure}
    \begin{subfigure}[b]{.24\textwidth}
    \includegraphics[width=0.9\textwidth]{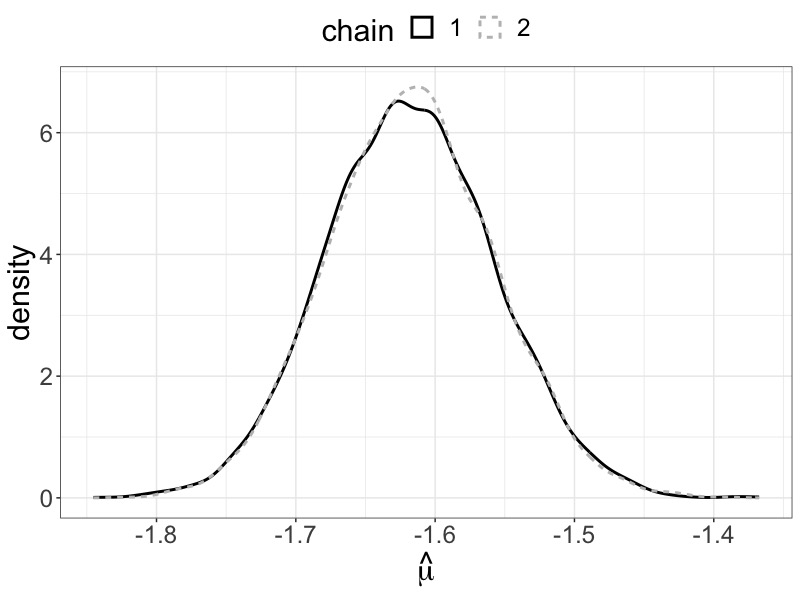}
    \caption{}
    \label{fig:mu_dens}
    \end{subfigure}
    \caption{Densities of posterior samples for the main parameters of interest of model RPT.}
    \label{fig:densities}
\end{figure}

\section{Estimated time-varying parameters of the RPT model on the real data application}

\begin{figure}[ht]
    \centering
    \includegraphics[width=0.8\textwidth]{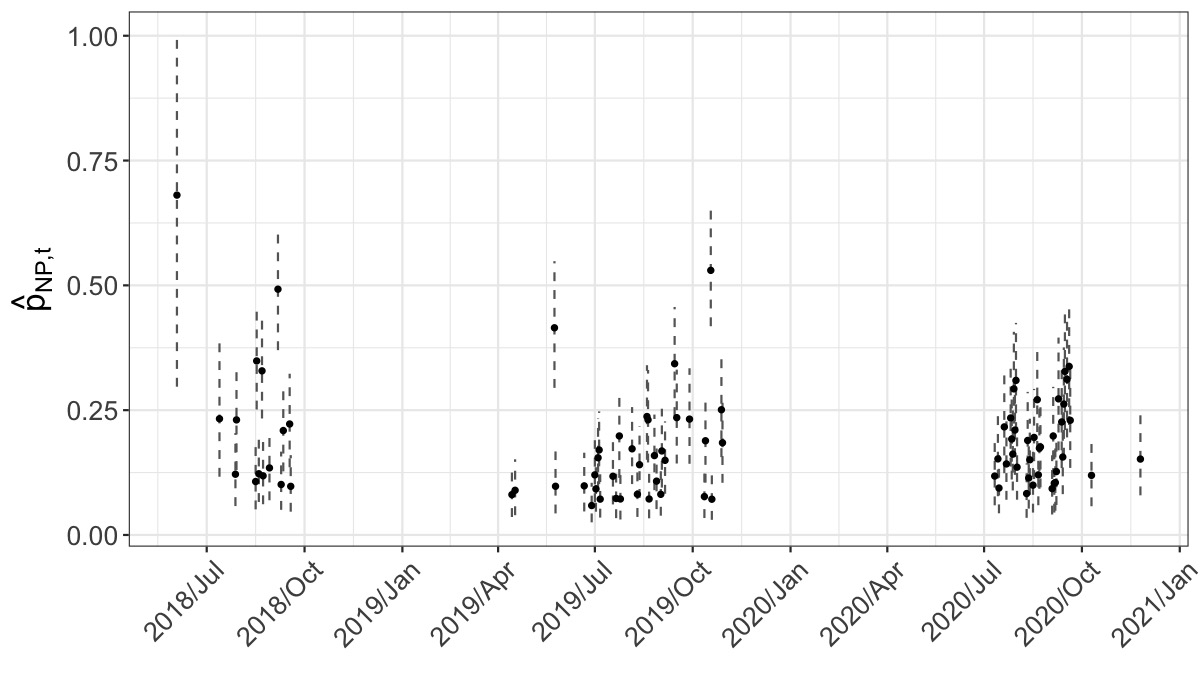}
    \caption{Posterior estimates and $95\%$ credible intervals for capture probabilities of resident and transient individuals at each sampling occasion.}
    \label{fig:pt_real}
\end{figure}

\begin{figure}
    \centering
    \begin{subfigure}[b]{.32\textwidth}
        \includegraphics[width=.95\textwidth]{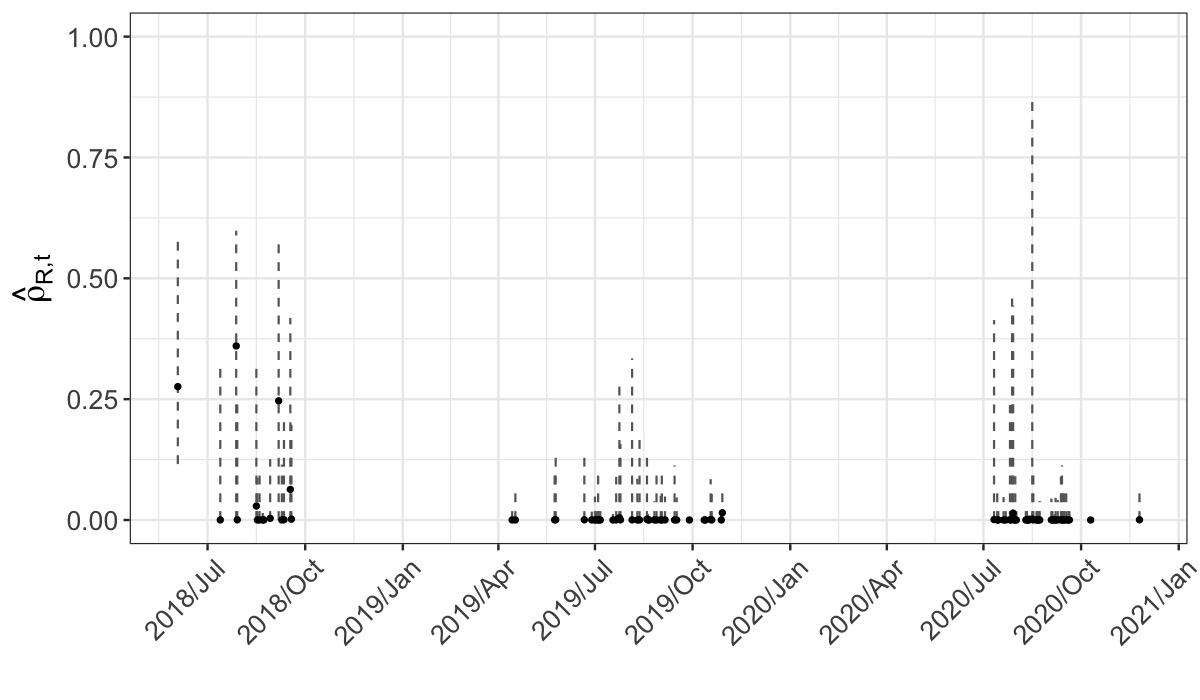}
        \caption{}
    \end{subfigure}
    \begin{subfigure}[b]{.32\textwidth}
        \includegraphics[width=.95\textwidth]{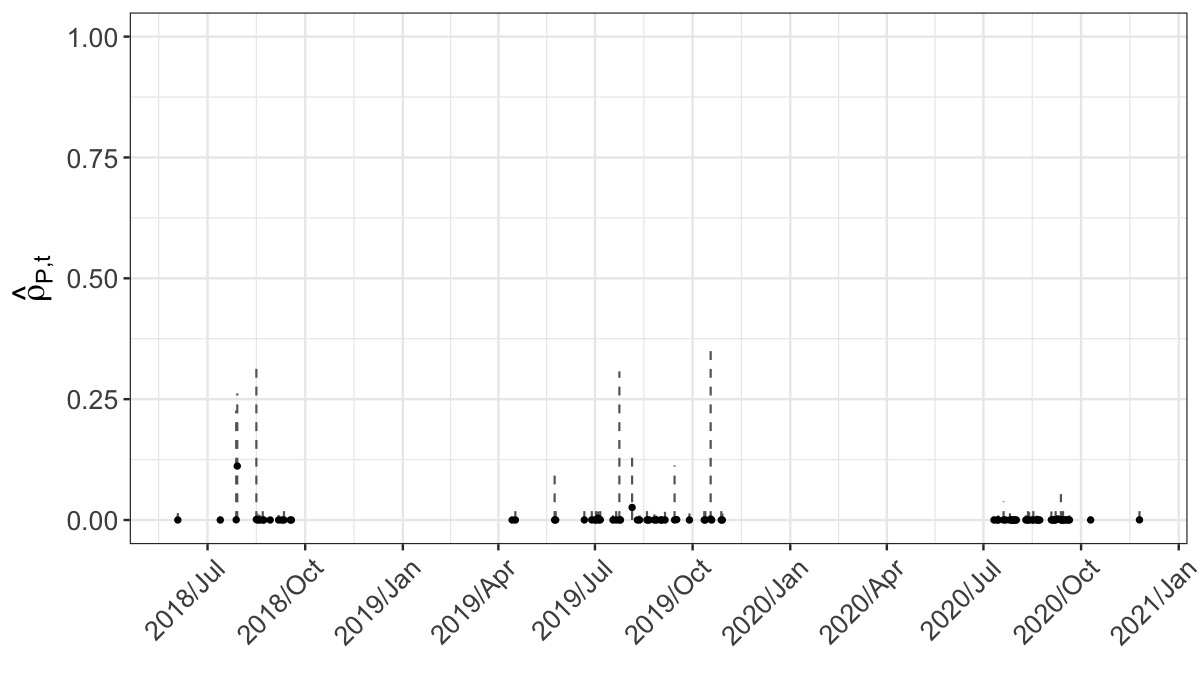}
        \caption{}
    \end{subfigure}
    \begin{subfigure}[b]{.32\textwidth}
        \includegraphics[width=.95\textwidth]{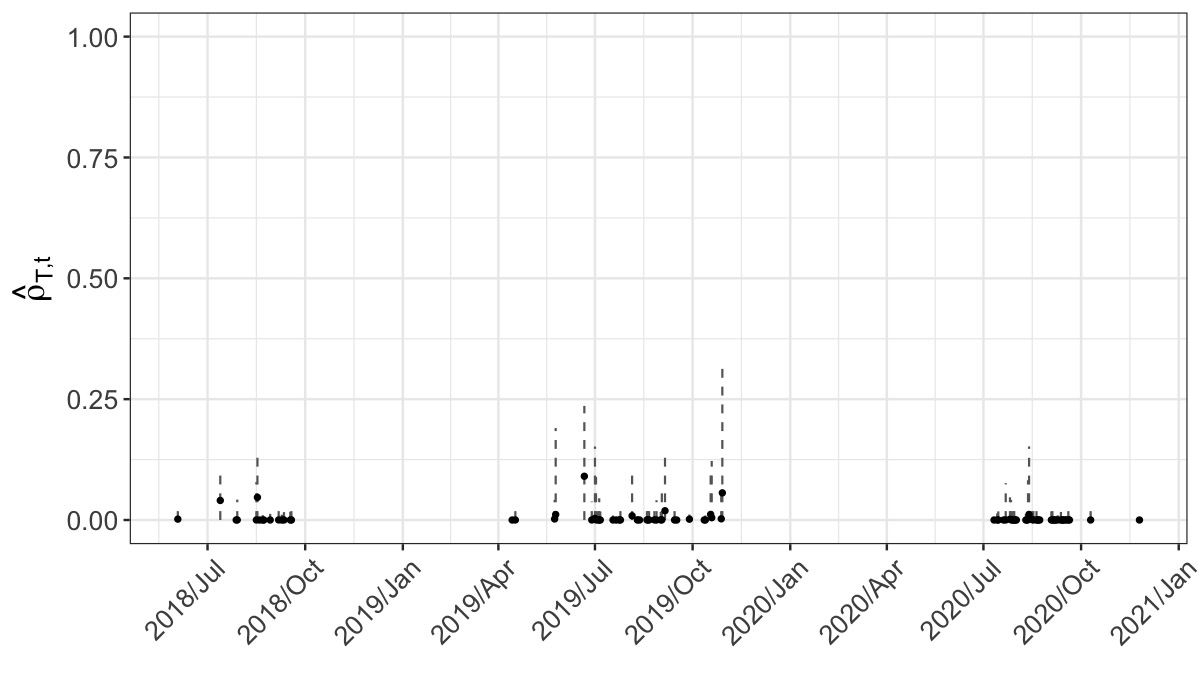}
        \caption{}
    \end{subfigure}
    \caption{Estimated recruitment probabilities (a) $\rho_{R,t}$, (b) $\rho_{P,t}$ and (c) $\rho_{T,t}$.}
    \label{fig:rhot_real}
\end{figure}

\end{document}